\documentclass{pnastwo}
\usepackage[dvips]{graphicx}
\usepackage{color}
\usepackage[dvips]{epsfig}
\usepackage{epsf}
\usepackage{epstopdf}
\usepackage{amssymb,amsfonts,amsmath}
\contributor{Submitted to Proceedings
of the National Academy of Sciences of the United States of America}
\url{www.pnas.org/cgi/doi/}
\copyrightyear{ }
\issuedate{Issue Date}
\volume{Volume}
\issuenumber{Issue Number}

\begin{document}
\title{Random Pinning Glass Model}

\author{Smarajit Karmakar\affil{1}{Departimento di Fisica, Universit\'a di Roma ``La Sapienza '',INFN,Sezione di Roma I, IPFC – CNR,
Piazzale Aldo Moro 2, I-00185, Roma, Italy}
\and Giorgio Parisi\affil{1}{}}

\maketitle 
\begin{article}
\begin{abstract}
Glass transition where viscosity of liquids increases dramatically upon decrease of temperature without any major change in 
structural properties, remains one of the most challenging problems in condensed matter physics \cite{09Cav, 11BB} in spite
of tremendous research efforts in last decades. 
On the other hand disordered freezing of spins in a magnetic materials with decreasing 
temperature, the so-called spin glass transition, is relatively better understood \cite{BookParisiSpin87, 05CC}. 
Previously found similarity between some spin glass models with 
the structural glasses \cite{87KT, 87KW, 87KWa, 99FP, 02MD} inspired development of theories of structural glasses 
\cite{RFOT, BAFRPA,MePa1,RFOT1, RFOT2} based on the scenario of spin glass transition. This scenario though looks very appealing is still far from being well 
established.
One of the main differences between standard spin systems to molecular systems is the absence of quenched disorder and the presence of translational invariance: it often assumed that this difference is not relevant, but this conjecture is still far from being  established.     The quantities, which are well defined and characterized for spin models, are not easily calculable for molecular glasses 
due to the lack of quenched disorder which breaks the translational invariance in the system and the characterization of the 
similarity between the spin and the structural glass transition remained an elusive subject still now. In this study we introduced 
a model structural glass with built in quenched disorder which alleviates this main difference between the spin and molecular 
glasses thereby helping us to compare these two systems: the possibility of producing a good thermalization at rather low temperatures is one of the advantages of this model.
\end{abstract}
\keywords{Glass transition| Random Pinning | Static Correlation Length | Replica Symmetry Breaking}

\dropcap{D}ramatic slowing down of relaxation process in almost all liquids when supercooled below the melting temperature still  
lacks a proper explanation \cite{09Cav, 11BB}. The increase of relaxation time in deep supercooled regime is so impressive that it 
becomes extremely difficult for the system to reach equilibrium in experimental time scales and eventually the liquid falls out of 
equilibrium with further decrease of temperature to under go a calorimetric glass transition. This transition is defined at a 
temperature where the viscosity of the liquids becomes $10^{13}$ poise \cite{88CAA, 92GS}. It is clear that this transition is ad hoc 
in nature and depends crucially on the choice of the parameter, but the main question, which remains to be answered, is whether there 
is a true thermodynamic transition below this calorimetric transition temperature. 

Many approaches to understand this remarkable slowing down in the dynamics of supercooled liquids invoke the existence of 
 a cooperative length scale \cite{00Ediger} associated with the collective rearrangements of particles and its divergence  
at the elusive glass transition. This scenario of glass transition is very similar in spirit to the critical phenomena seen in the continuous
phase transition. The slowing down is believed to be caused by the difficulty to rearrange sets of ever increasing number of 
particles in a collective fashion with decreasing temperature or increasing density. Attempts to estimate this cooperative length scale 
remained one of the major difficulties due to the lack of identification of an order parameter characterizing the structural glass 
transition. This is primarily due to lack of growth of an obvious order in the system with decreasing temperature. 

Progresses made in recent years to identify such a length scale is really encouraging. Dynamic heterogeneity length 
scale from the analysis of 4-point density-density correlation function \cite{98P,05Berthier, 06BBMR, 09KDS}, Point-to-set length scale 
\cite{08BBCGV}, patch length scale\cite{09KL}, length scales associated with non affine displacement of particles \cite{10MGIO},  
from finite size scaling of configurational entropy \cite{09KDS} and density of states \cite{12KLP} are a few to name. 
Unfortunately there is still no general consensus about the importance of these length scales to glass transition and their relation 
to each other. The main hurdle in reaching such a 
goal is that these length scales are only accessible in computer simulation studies because of the requirements of microscopic
details to compute them. So these length scales can only be estimated in small parameter range where it grows very modestly thereby making 
it almost impossible to see the divergence while approaching the glass transition. 

Based on these ideas of growing length scale and remarkable similarity seen in the dynamics  of $p$-spin glassy model with the structural 
glasses, Kirkpatrick, Thirumalai and Wolynes \cite{RFOT, RFOT1} proposed the Random First 
Order theory (RFOT) of glass transition. This theory is very much in spirit with the Adam-Gibbs Theory \cite{65AG} proposed much earlier. 
This theory seems to suggest that  $p$-spin glass models and structural glasses should belong  the same universality class. 
 However there is an inherent difference between these two models, that is the existence of quenched disorder in spin glass and not in the 
structural glass models. Lack of quenched disorder in structural glasses makes it rather difficult to calculate the quantities like spin-glass 
type order parameter and susceptibility: moreover the whole low temperature phase can not be accessed by simulations. In a recent mean field and 
renormalization group study \cite{12CB}, it was proposed that with random pinning one can explore the ideal glass phase as in the 
temperature - concentration of frozen particles plane there exists a critical point where relaxation time does not diverge while going 
from liquid to glass phase. Similar studies done in the Mode Coupling Theory frame work \cite{11Krakoviack, 12Szamel} also confirm this
picture. All these studies and some other recent studies \cite{03Kim, 11KMS, 12KLP, 12BK,12MP} clearly show that exploring the glassy state 
in the random pinning geometry can be very insightful with the added advantage of built in quenched disorder. This can enable us to do 
a replica theoretic calculations for these kind of particle models to shed more light on the soundness  of the replica approach to  structural glasses.  

\begin{figure*}
\begin{center}
\epsfig{file=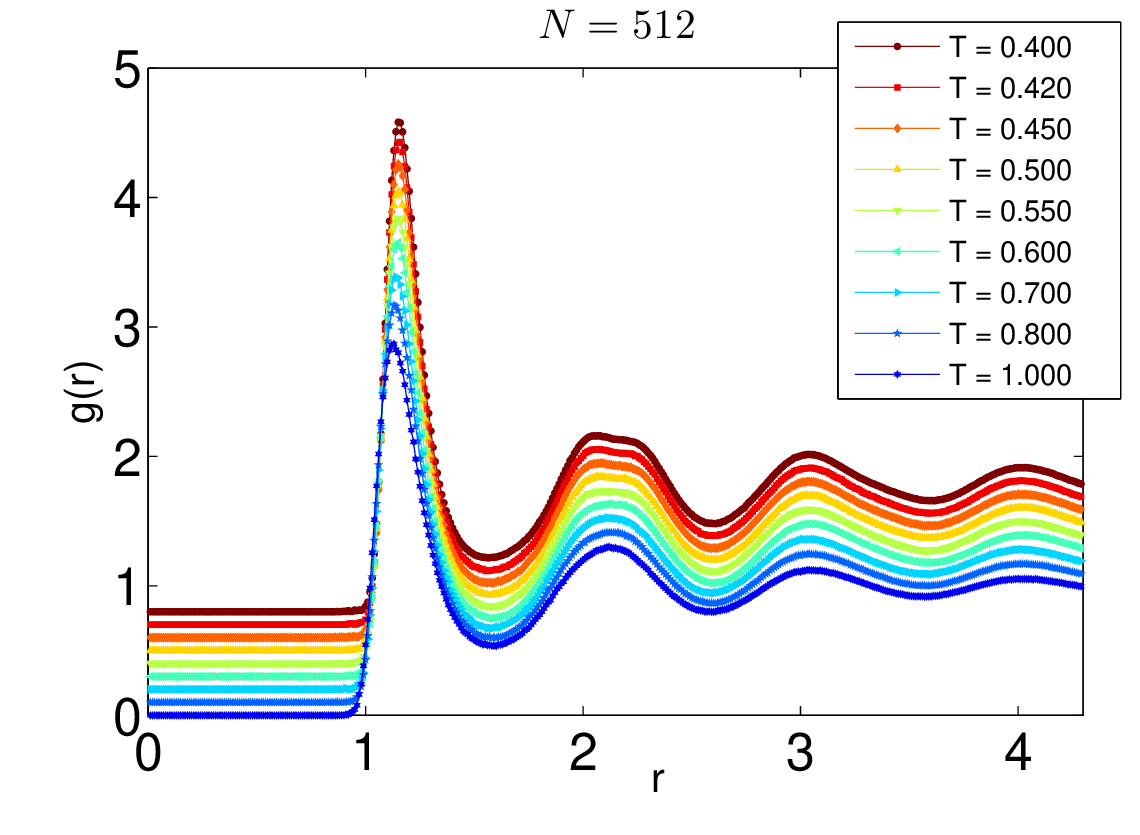,scale=0.40,angle=0,clip=} 
\epsfig{file=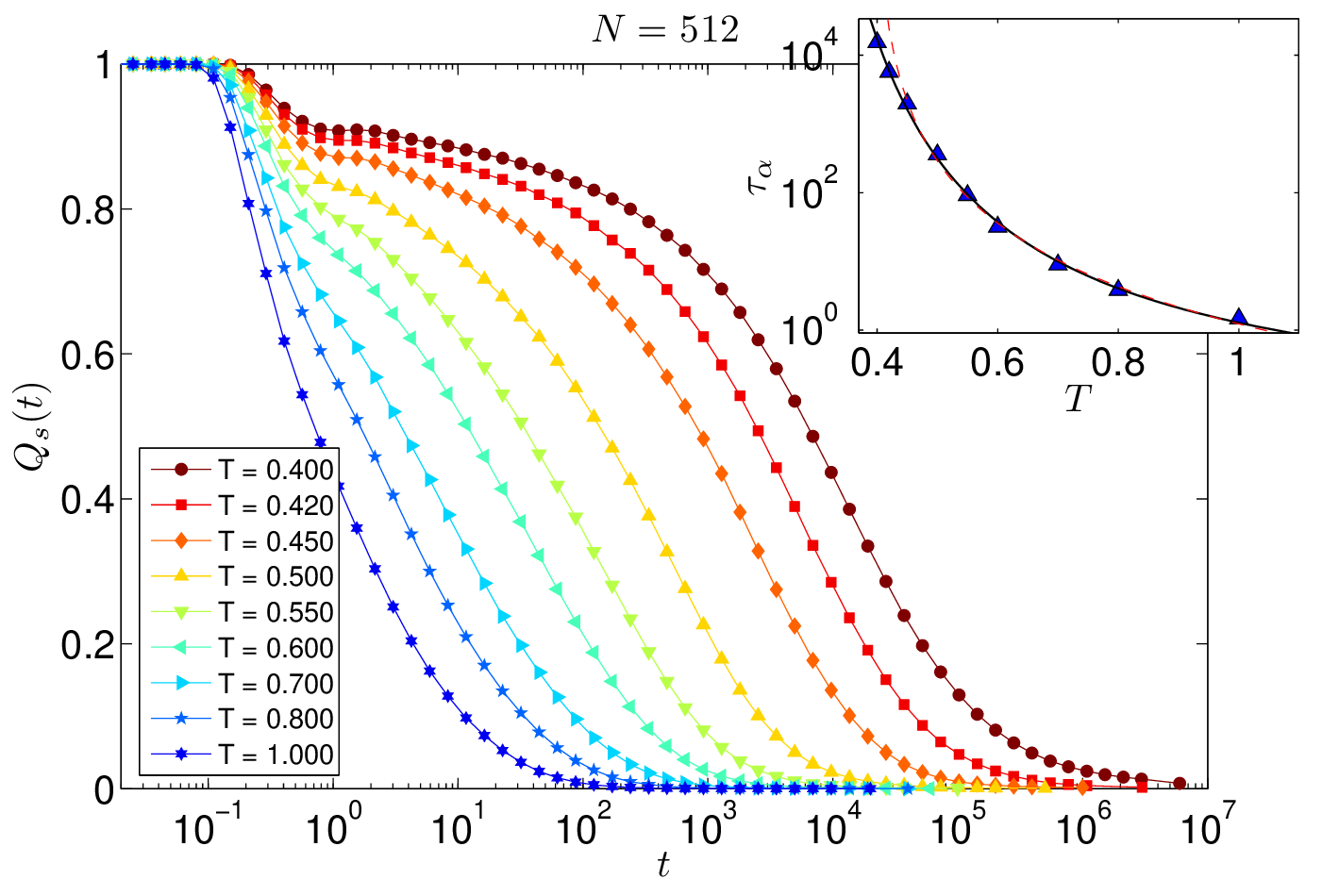,scale=0.35,angle=0,clip=} 

\epsfig{file=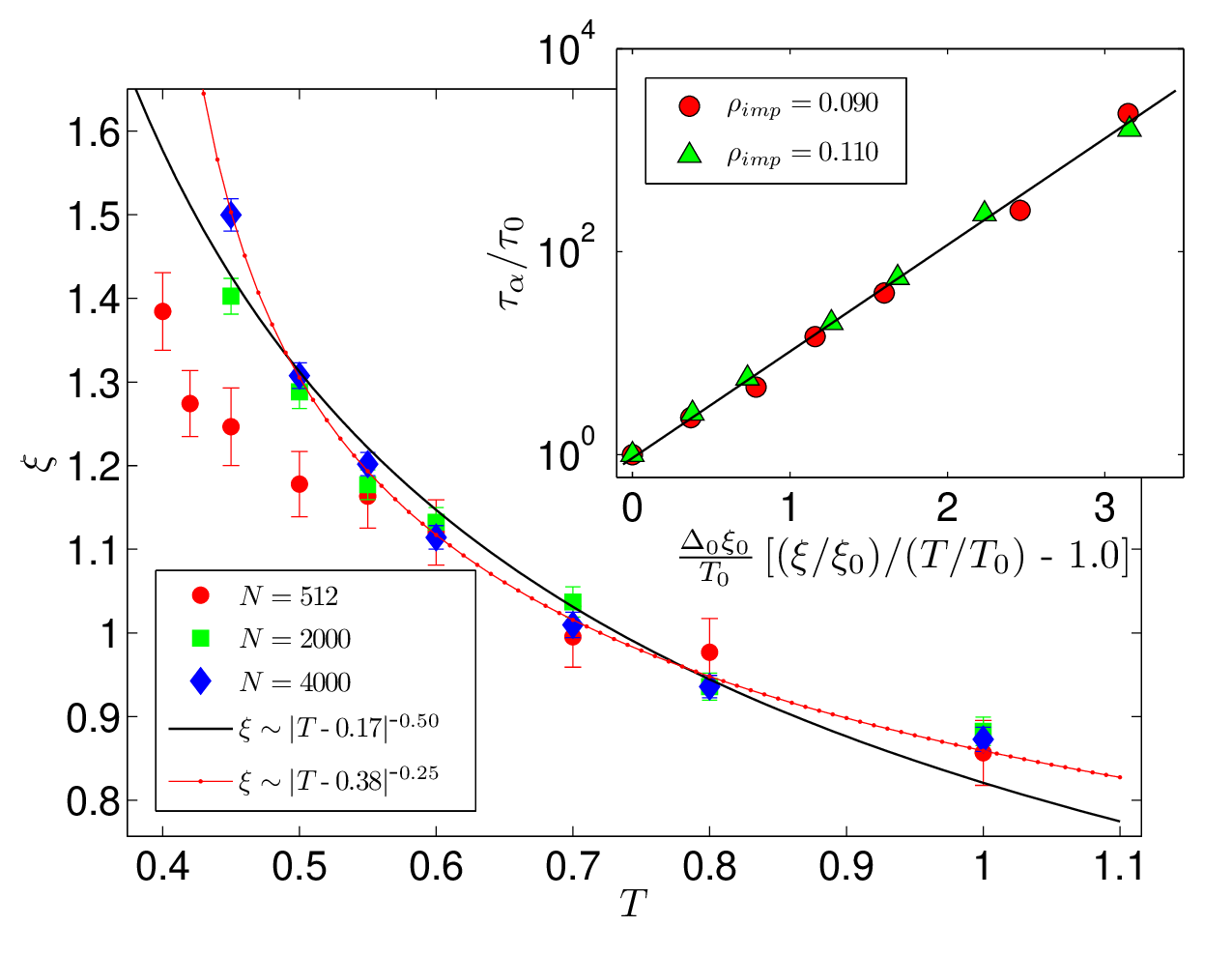,scale=0.35,angle=0,clip=} 
\epsfig{file=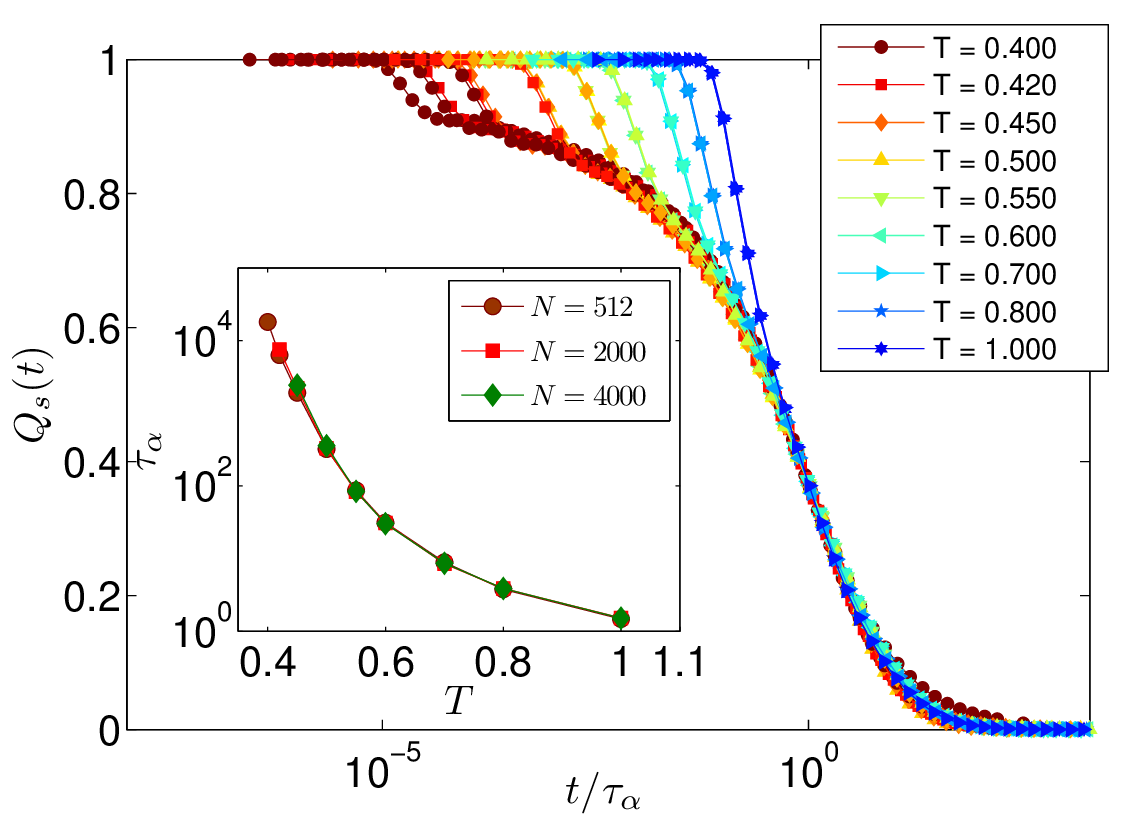,scale=0.41,angle=0,clip=} 
\caption{ Top Left Panel: Pair correlation function $g(r)$ for $N = 512$ system size for different temperatures. The curves for different 
temperatures are shifted vertically for clarity. Top Right Panel : Self part of the two point correlation function $Q_s(t)$ for different 
temperatures. The inset shows that $\alpha$-relaxation time $\tau_{\alpha}$ calculated as the time when $Q_s(t)$ goes to $1/e$ of its 
initial value. The line is the fit to the data using Vogel-Fulcher-Tamman Formula (VFT) with $T_K \simeq 0.17$. The dynamic transition 
temperature (MCT) is estimated to be $T_d \simeq 0.394$ ( red dashed line ). Bottom Left Panel : The extracted static length scale from 
the replica overlap correlation functions ( see the method section for details ). The red circles, green squares and blue diamonds are 
from the system with $N = 512$, $N = 2000$ and $N = 4000$ respectively. The estimation of the length scale from  $N = 512$ system size 
seems to have some finite size effects which almost disappear for $N = 2000$ and $N = 4000$ system sizes. The black line is the fit 
to the data using the form $\xi(T) \sim |T - T_K|^{\gamma}$, with chosen $T_K \simeq 0.17$ and $\gamma \simeq 0.50$ and the red 
line with dot is the same fitting where we allowed all the parameters to vary and the resulting $T_K \simeq 0.38$ and 
$\gamma \simeq 0.25$. Inset shows the dependence of the 
relaxation time with the length scale for $N = 4000$ system sizes. One can see some degree of universality in the relation between 
relaxation time and length scale (see text for details). Bottom Right Panel: Time-Temperature superposition for the $Q_s(t)$ for all 
system sizes $N = 512, 2000, 4000$ for all the studied temperatures. The very nice collapse of the data confirms that these model 
indeed has all the usual features of a glassy system. Inset shows the temperature dependence of $\tau_{\alpha}$ for different system sizes.}.  
\end{center}
\label{dynamicalDetails}
\end{figure*}

In general a system of particles of equal size interacting via a radially isotropic pairwise potential will have liquid to crystal phase 
transition with decreasing temperature for dimensions $d\le3$. At higher dimensions it was shown that crystallization is strongly 
suppressed \cite{11CIPZ}. So to form a glass it is very important at least for dimension smaller than $4$ to introduce frustration in 
the system to prevent it from quickly falling in to the crystalline global minimum. In spin glasses the random interactions between 
the spins are the source of this frustration and in structural glasses compositional disorder, for example different sizes of the 
particles or the asymmetric interaction between different types of particles, usually generates the required frustration. 
Due to presence of these extra degrees of freedom in general glass models they are often very hard to equilibrate at lower temperature 
as one needs also to equilibrate these extra degrees of freedom. It would be nice to have a glassy model system without extra degrees of freedom. In this article we propose to generate
the required frustrations by random pinning:  we have study a system where 
we do not have any randomness in the interaction potential nor do we have any compositional disorder. Our model system consists of 
particles of equal size with some fraction $\rho_{imp}$ of them frozen randomly in space. If sufficient number of these 
particles is frozen randomly in space at some high temperature there is enough frustrations in the system 
to force the system to remain in disorder state. As there is no other degrees of freedom to equilibrate apart from the positions of the 
particles we expect this model can be equilibrated to lower temperatures than the usual models to study the lower temperature phase 
of the supercooled liquid. With the quenched disorder, this model is also very attractive to compare it with the spin glass models. At high density of quenched particles no crystallization is present, however the glass phase is also absent: we will show that there is an intermediate region with a small, but not too small fraction of quenched particles, where a part of the usual structural glass phenomenology survives and the system does not crystallize, also at very low temperatures. Our construction differs from that of \cite{12CB} as far as the quenched disorder is crucial to avoid crystallization and finding a phase transition.

\noindent{\bf Results :}
It turns out that one needs to freeze around $\rho_{imp} = 9\%$ of particles to get a system which will not show crystallization at 
least for bigger system sizes ($N > 250$). For smaller systems this amount of frozen particles is found to be not sufficient to prevent
the crystallization. With $\rho_{imp} = 11\%$, even the smaller system sizes do not show any tendency to crystallize. So for the present 
study we choose to work with $\rho_{imp} = 11\%$ for all the system sizes studied. We restricted ourselves to small system sizes mainly 
because we wanted to achieve full equilibration of the system to 
very low temperatures. The studied system sizes are in the range $N\in \{100, 4000\}$. In the top panel of Fig.\ref{dynamicalDetails}, we
showed the pair correlation function for different temperatures to confirm that there is no sign of incipient crystallization in this temperature 
range. In top right panel we showed the self part of the average two point density correlation function 
$\langle Q_s(t)\rangle$ (see Dynamics in the Method Section for details) for $N = 512$ system size for different temperatures
in the range $T \in \{0.400, 1.000 \}$. One can see the nice development of the plateau in the correlation function with 
decreasing temperature. In the inset of this panel, we have shown the temperature dependence of the $\alpha$-relaxation 
time and the line is the fit to the data using Vogel-Fulcher-Tamman (VFT) formula with the estimated divergence temperature 
$T_K \sim 0.17$ for $N = 512$ system size. The dynamic transition temperature or the mode coupling cross over temperature is 
estimated to be $T_d \sim 0.394$ by a power fit. One should keep in mind that these extrapolated estimation of the divergence temperatures may not be very reliable as the range of the data is not very big.  

As our model is not translationally invariant due to the presence of the quenched disorder defining overlap between two replicas 
becomes easy and unambiguous. For example if we have a system with translational symmetry then while defining the overlap between two 
replicas we need to take in to account the fact that two replicas can be similar even though they may be translated or rotated in space \cite{98CoPa}. 
We defined the overlap between two replicas using the window function $w(x)$ which is $1$ if $x<0.30$ and zero otherwise. We say the two
replicas are similar if for each particle at position $\vec{r}$ in replica $1$ there is a particle of replica $2$ within a sphere of 
radius $0.30$ (see Replica overlap in Methods Section for details). We also defined local overlap 
and calculated the corresponding spatial correlation function to extract the static length scale over which the replicas 
are similar. In the bottom left panel of Fig.\ref{dynamicalDetails} the extracted static correlation length of two replicas
(see the Replica Overlap and Extraction of static length scale in Method section for details \cite{JANUS}). The finite size effect seems to die out 
quickly once we go to $N = 2000$ system size for lower temperatures. The modest growth of the static length scale in this temperature 
range is very similar to other glass models. The inset shows the dependence of the relaxation time with this length scale 
(see the relation between length scale and relaxation time in Method section for details). The apparent universality for two different 
frozen particles density is also in agreement with the recent findings for usual glass models \cite{12HMR, 12KP}. 

\begin{figure*}[!h]
\begin{center}
\includegraphics[scale = 0.40]{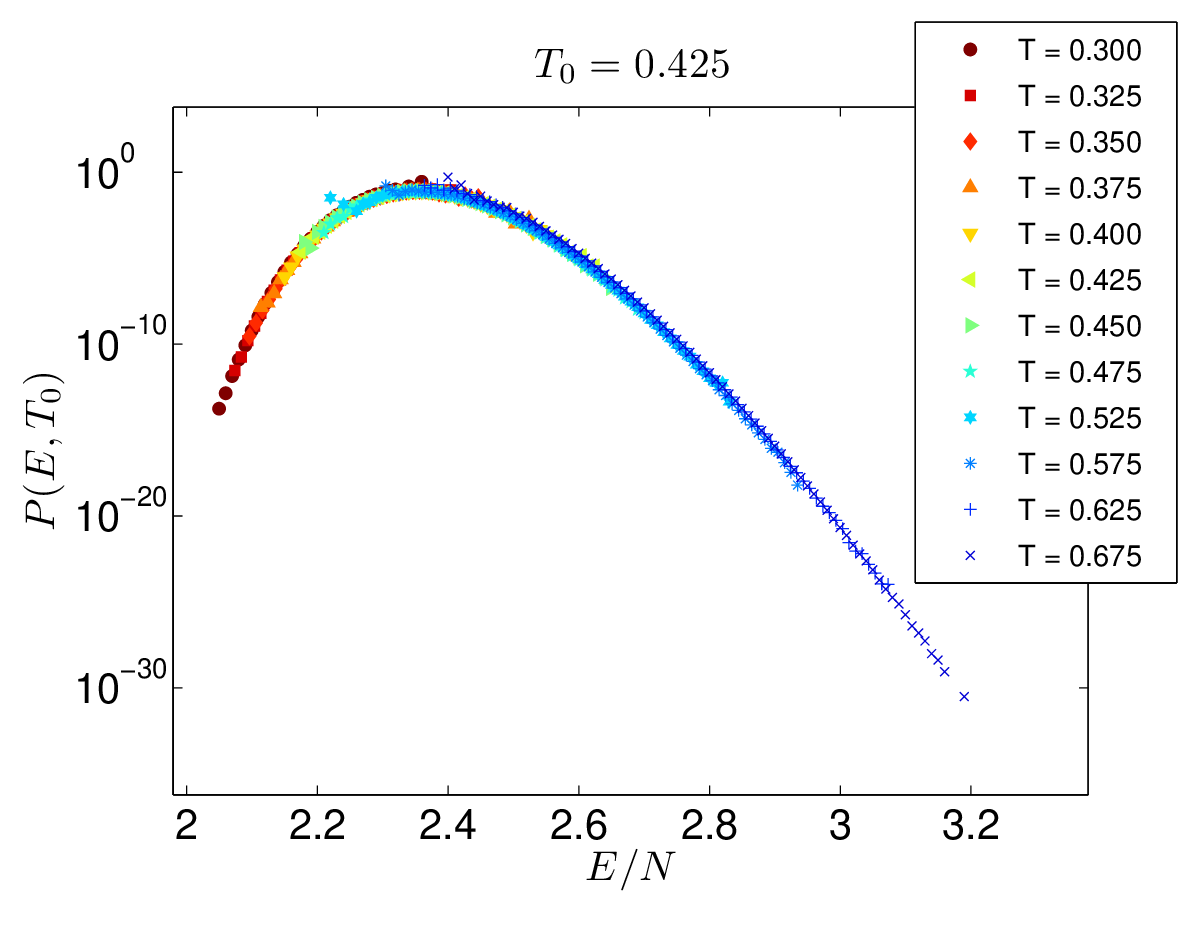}
\includegraphics[scale = 0.42]{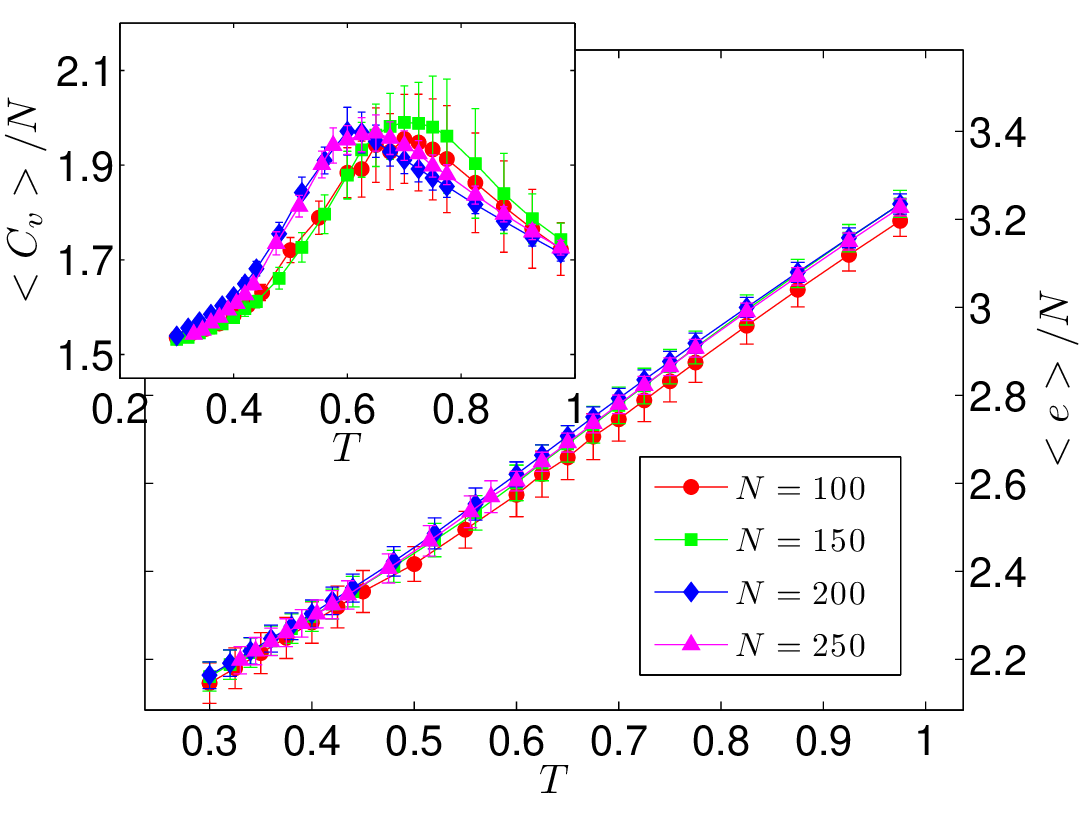}
\caption{ Left Panel: The collapse of distribution of energy $P(E)$ for different temperature on the distribution of $P(E,T_0)$ with 
$T_0 = 0.425$ in the Parallel tempering run for $N = 100$
system size according to the Eq.\ref{reweighting} to check the equilibration. The nice collapse confirms that the very good equilibration 
is achieved using the Parallel tempering method for lower temperatures. Right Panel: The temperature dependence of the energy for different
system sizes. One can see that the finite size effect is not very strong here. Inset shows the specific heat calculated from the fluctuation
of potential energy for different system size. Some finite size effect can be seen here.}
\label{parllTmp}
\end{center}
\end{figure*}

To achieve equilibration at further lower temperature we implemented Parallel Tempering simulation methods following 
\cite{00YK}, but restricted ourselves to systems up to $N = 250$. The details of the simulation method and the parameters
used are given in Parallel Tempering methods parts in Method Sections. We parallelized the Parallel Tempering method 
using Message Passing Interface (MPI) routines to speed up the simulation. We run different replica in different computer cores
and we found that the system can be equilibrated to very low temperature within reasonable CPU time (around $12$ hours for $N = 250$ 
particles using $16$ replicas in $16$ cores). We checked that system does not crystallize even at the lowest temperature studied. 
We believe parallel tempering methods works so well for our model as it has only positions to equilibrate and does not have any other 
compositional disorder which needs further equilibration. In left panel of Fig.\ref{parllTmp}, we have collapsed the probability 
distribution of potential energy $P(E,T)$ for different temperatures according to the ansatz Eq.\ref{reweighting} on the probability
distribution of the reference temperature $T_0 = 0.425$, to ascertain whether proper equilibration is achieved in our simulations (see Parallel 
tempering part in Method section for details). The nice collapse of the data indicates that the equilibration is achieved to a very good 
accuracy. The right panel of Fig.\ref{parllTmp} shows the temperature dependence of the average potential energy and the inset shows the 
corresponding specific heat calculated from the fluctuation of potential energy. Although average potential energy does not show strong
finite size effects one can see somewhat strong finite size effect in the specific heat. The nicely developed peak in the specific heat 
seems to be a precursor to a possible second order phase transition as also seen in \cite{98CCP}. The usual discontinuity in the specific heat seems to remain rounded in a volume independent faction. Notice that at the lowest temperatures we have simulated the specific heat drops to the DulongÐPetit value, suggesting that harmonic degrees of freedom are the mostly relevant ones at these temperatures.

\begin{figure*}
\begin{center}
\includegraphics[scale = 0.31]{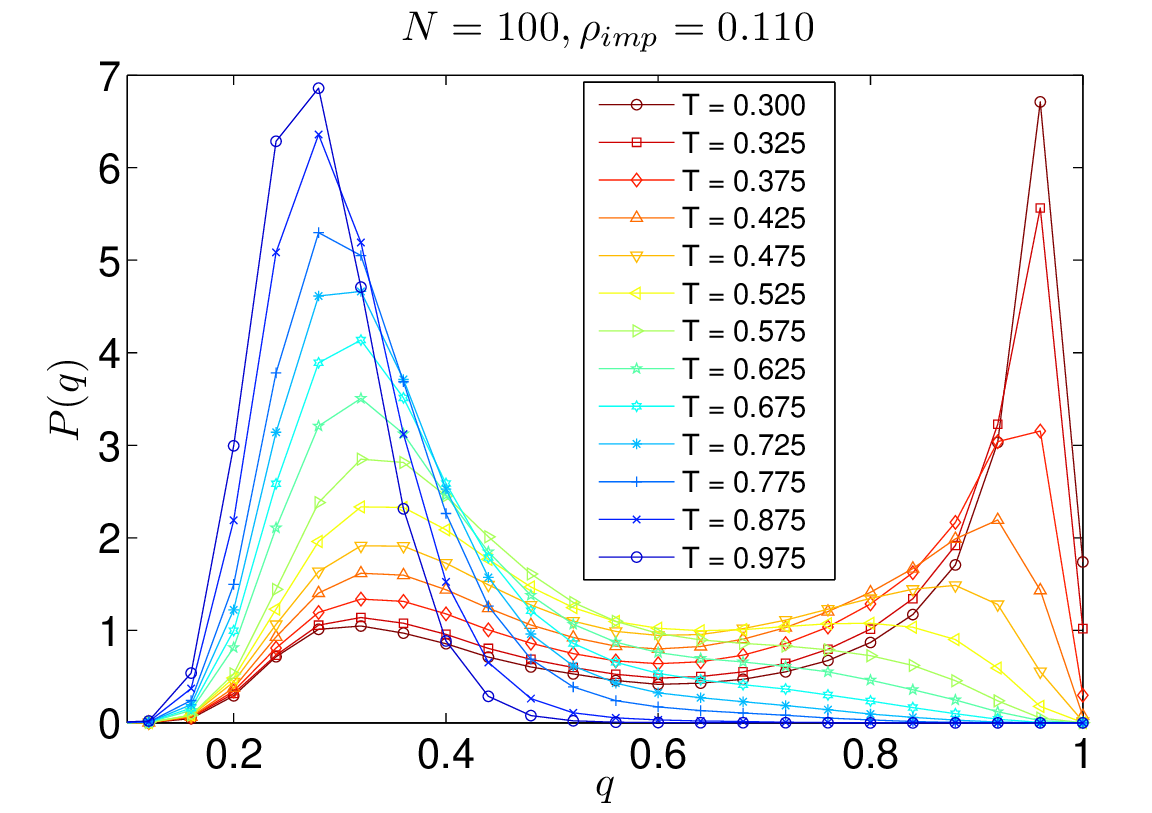}
\includegraphics[scale = 0.29]{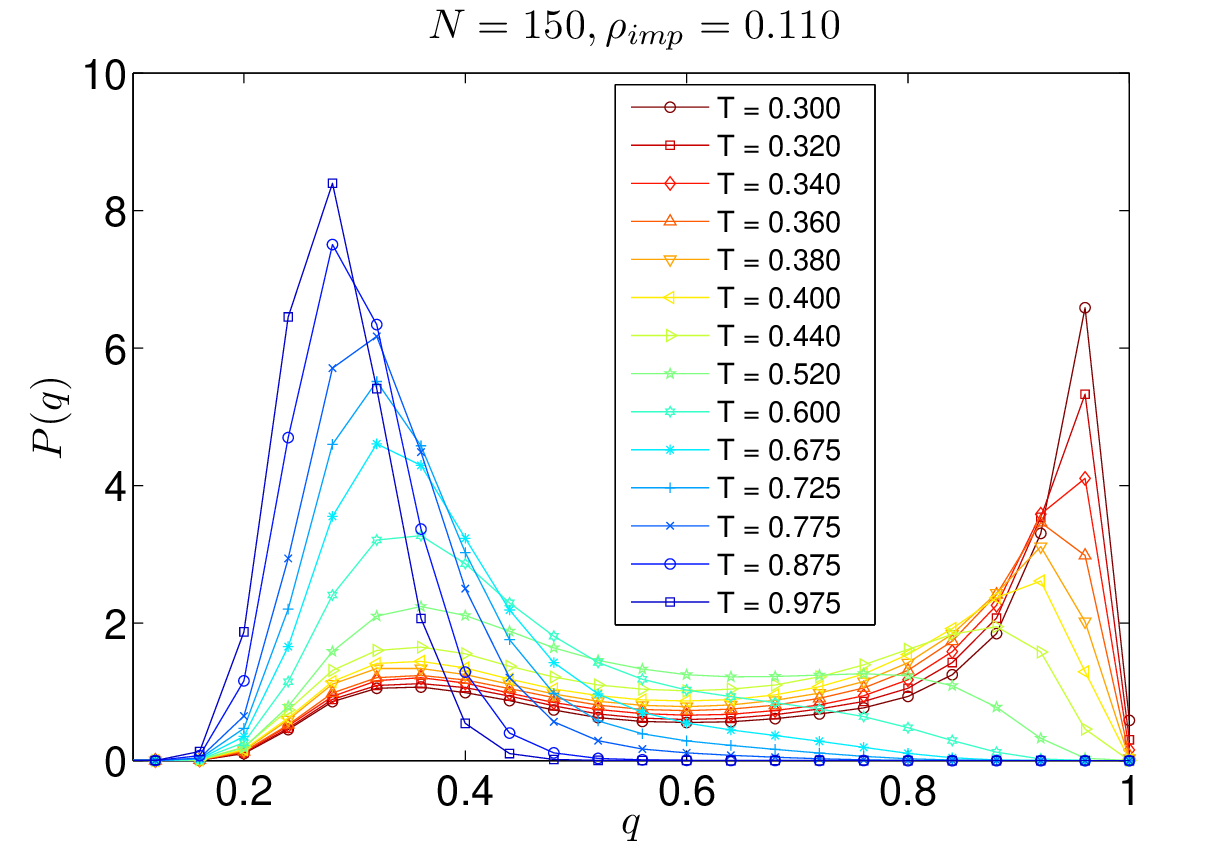}
\includegraphics[scale = 0.30]{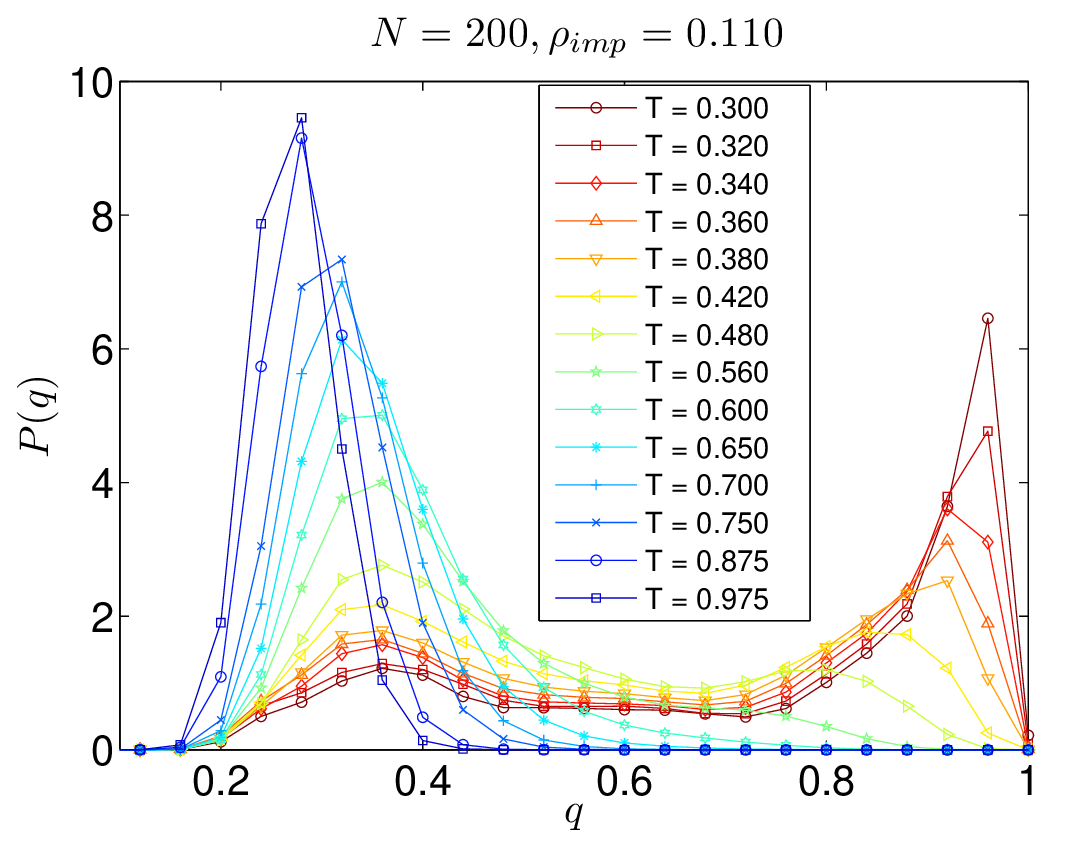}

\includegraphics[scale = 0.32]{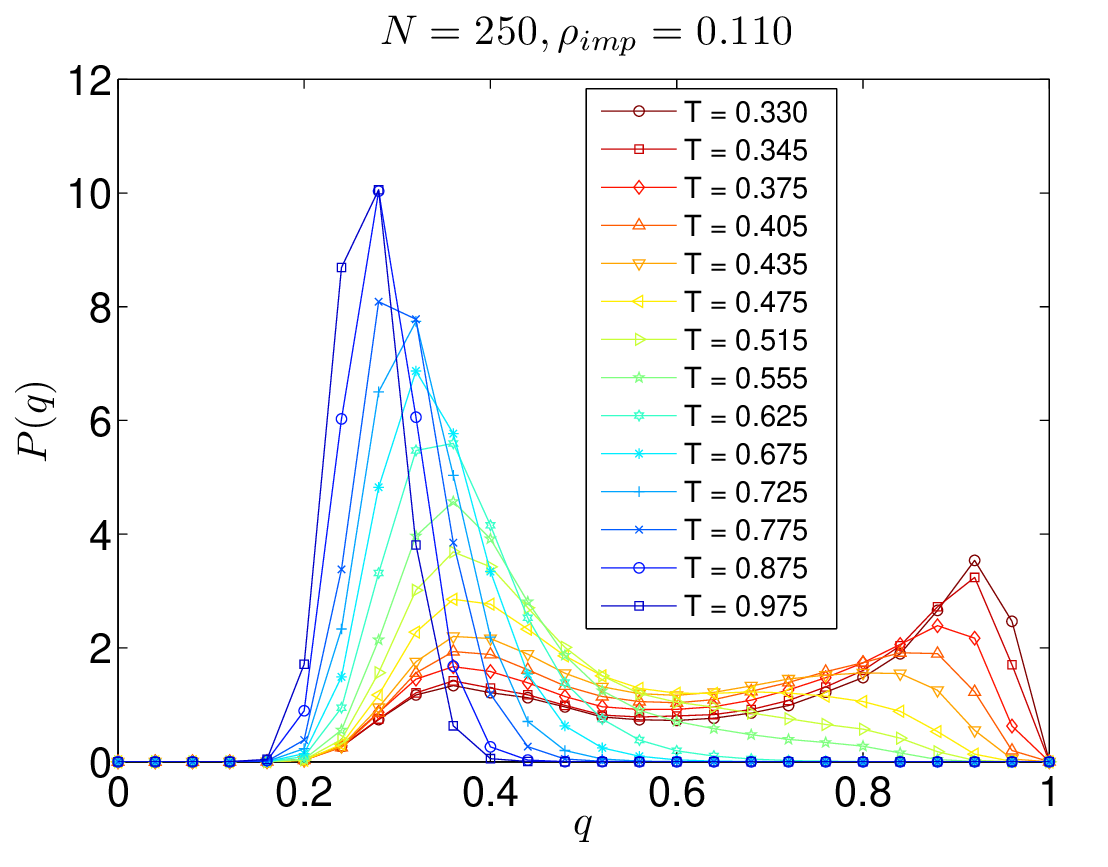}
\includegraphics[scale = 0.33]{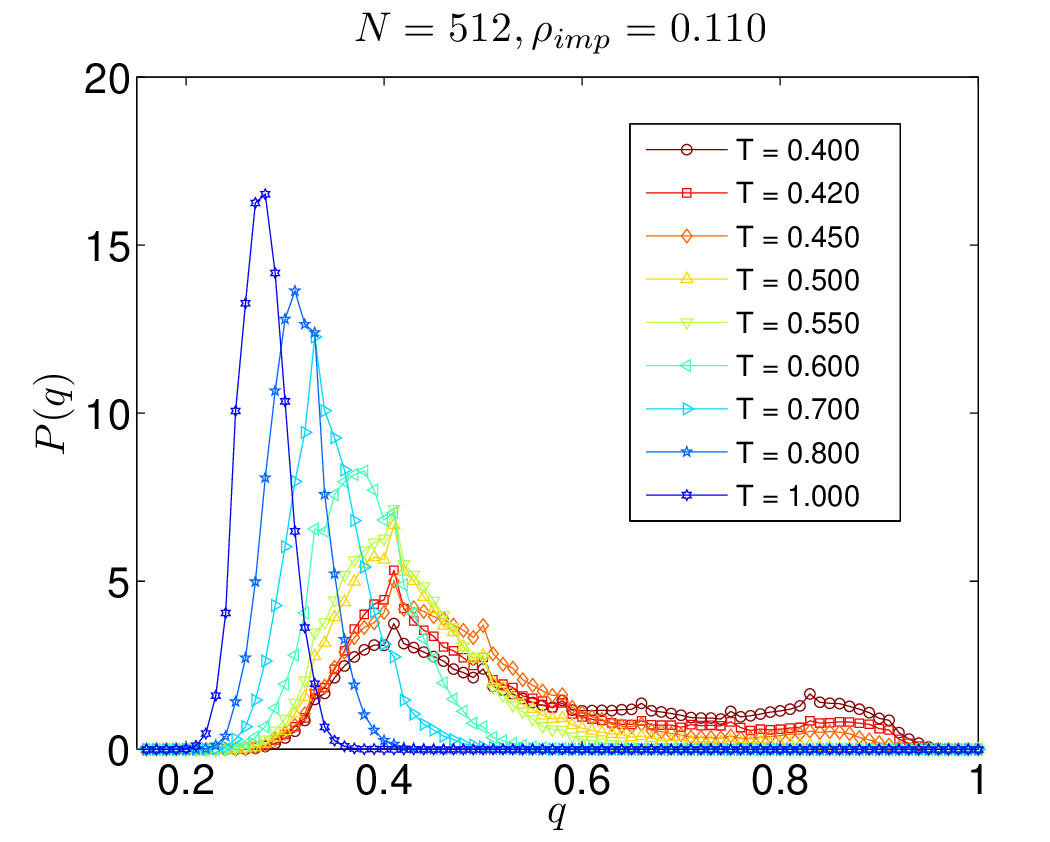}
\includegraphics[scale = 0.32]{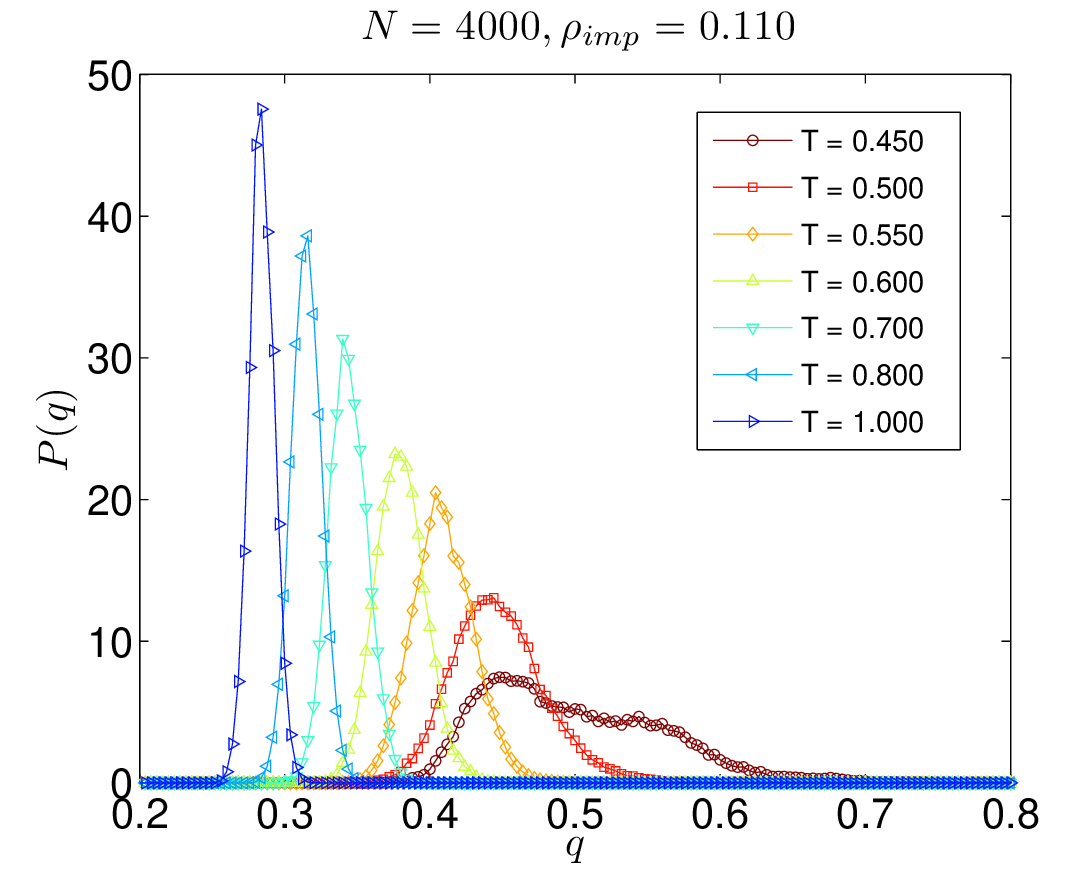}
\caption{ Temperature evolution of the Probability distribution of overlap $q$ for six different system sizes $N = 100, 150, 200, 
250, 512, 4000$.}
\label{probDistq}
\end{center}
\end{figure*}

In Fig. \ref{probDistq}, the temperature evolution of the distribution of overlap $q$ is shown for six different system sizes. One can 
clearly see the change of the distribution from Gaussian to Bimodal with decreasing temperature. The distribution seems to deviate 
from the Gaussian one at temperature close to the temperature where the specific heat also shows peak as a function of temperature for 
that system size. The shape of the distribution as function of the temperature recalls what happens in mean field theory in the replica approach: below the critical temperature a peak at higher values of $q$ appears while the low $q$ peak has an intensity that is proportional to the temperature.

\begin{figure*}
\begin{center}
\epsfig{file=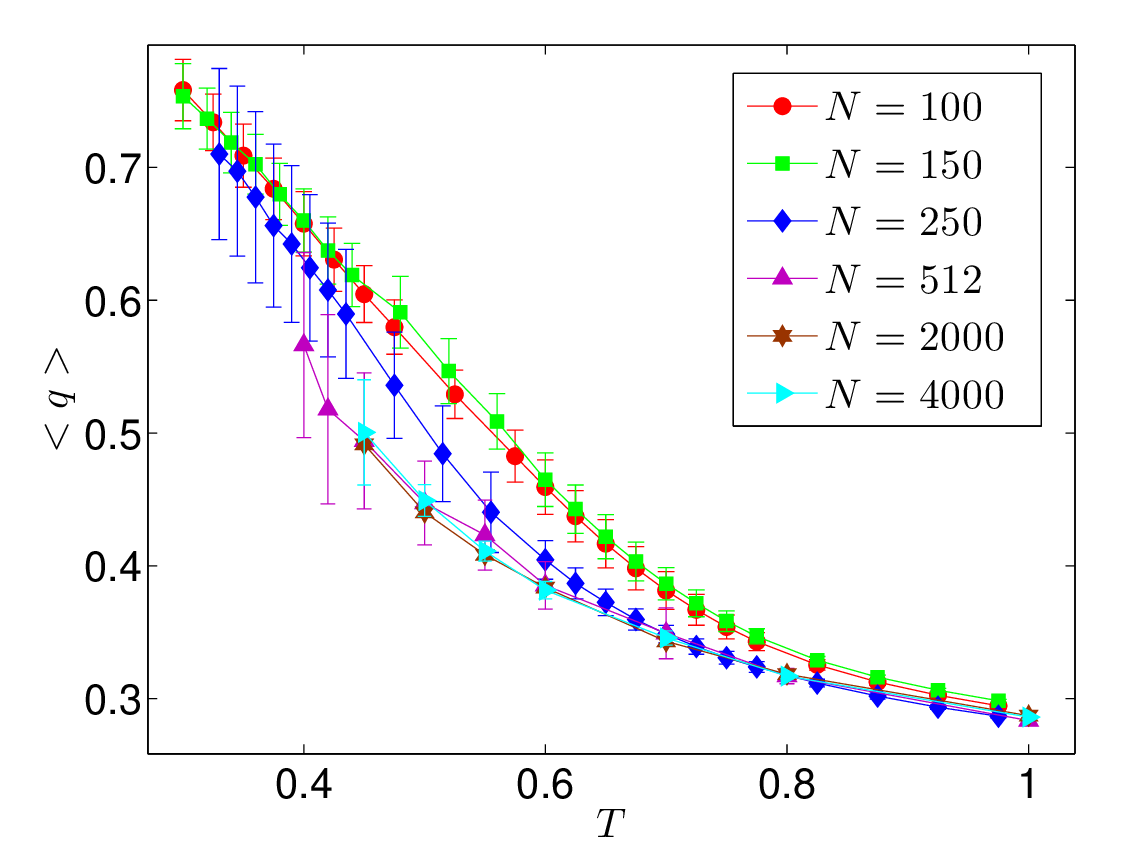,scale=0.30,angle=0,clip=} 
\epsfig{file=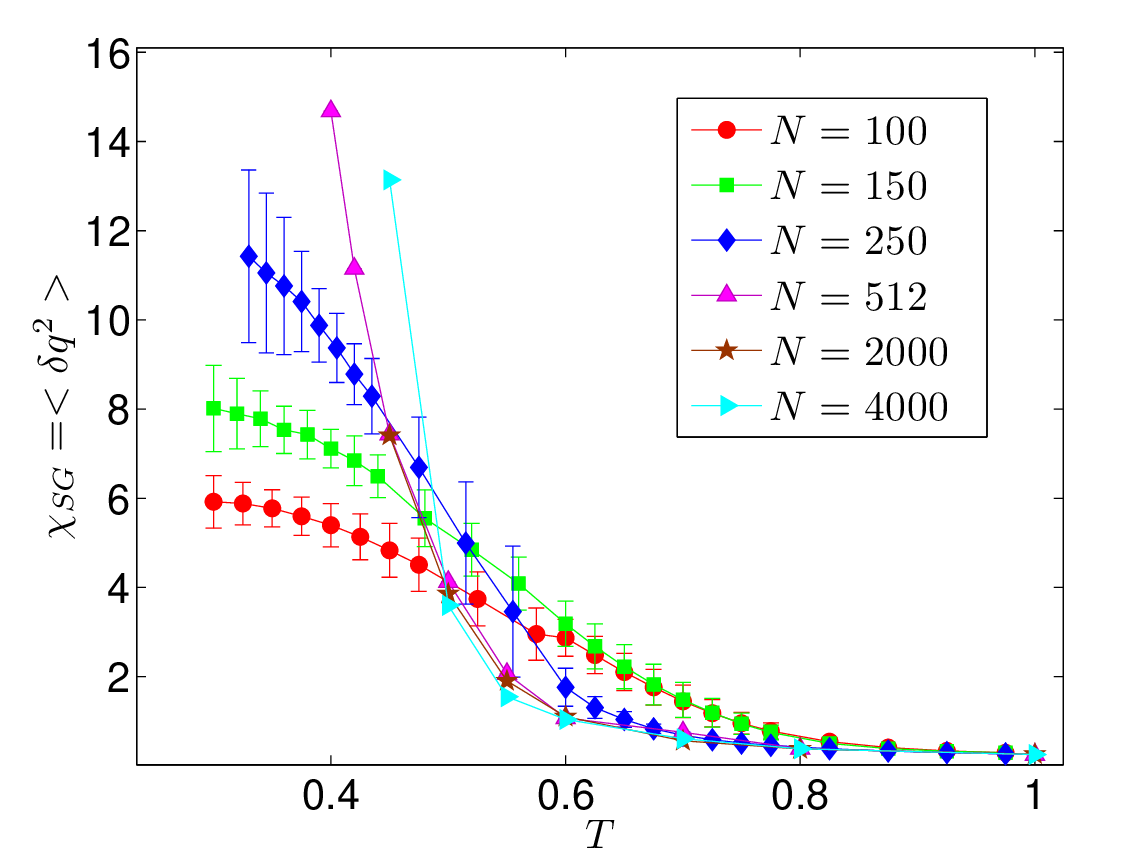,scale=0.30,angle=0,clip=} 
\epsfig{file=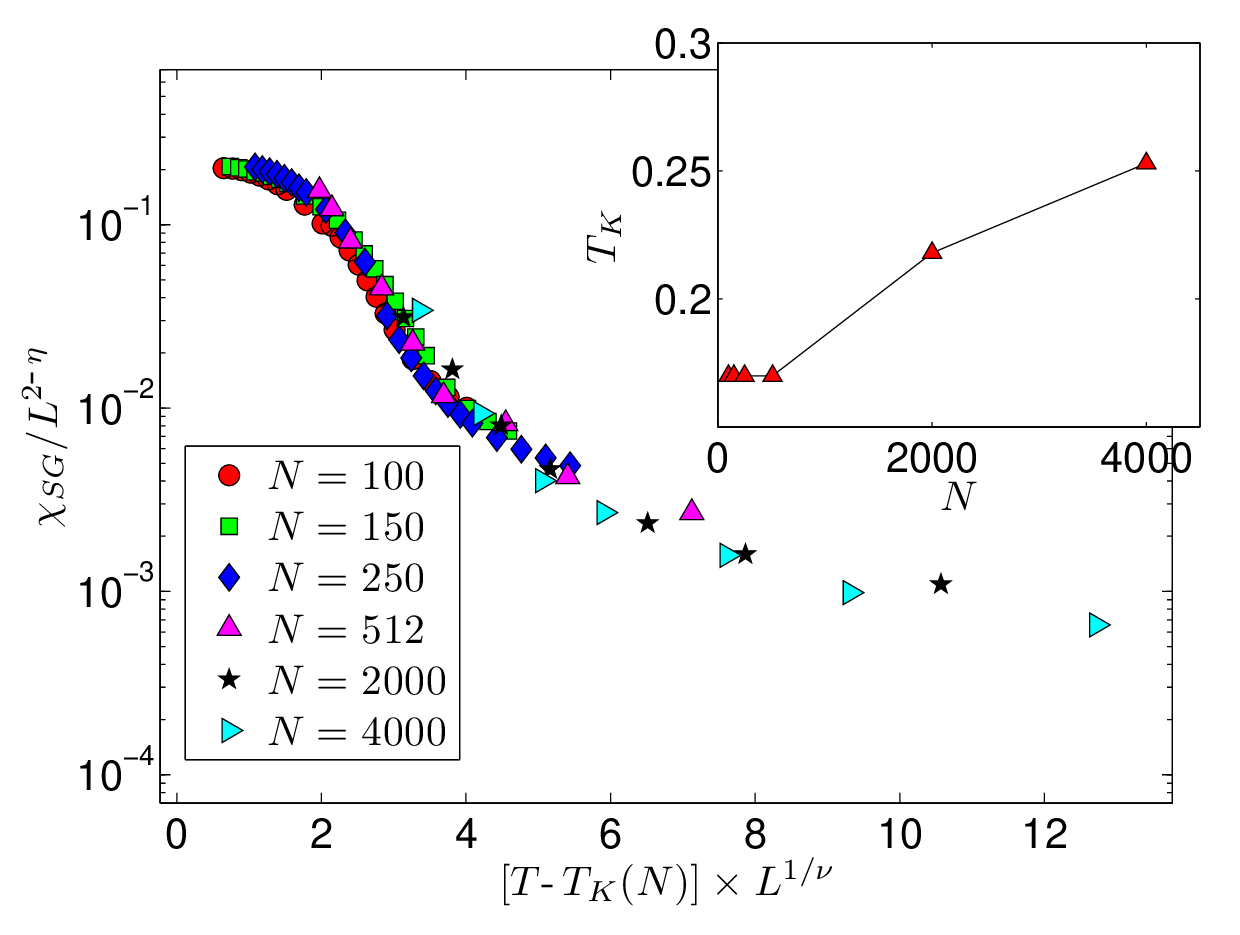,scale=0.28,angle=0,clip=}
\caption{ Left Panel : Temperature dependence of overlap for different system sizes. Bigger system size seems to show sharper change of 
overlap $q$ with temperature. Middle Panel: The spin glass susceptibility is plotted as a function of temperature for the studied system 
sizes. The strong variation with temperature is really remarkable. Notice the strong system size dependence of susceptibility at lower 
temperature. Currently we can not say whether this susceptibility will diverge with decreasing temperature in thermodynamic limit as 
the system sizes studied here are still quite small. One needs to do bigger system sizes to find out the possible divergence of the spin
glass susceptibility at lower temperature. Right Panel: Scaling collapse of the spin glass susceptibility using the scaling ansatz 
Eq.\ref{chiAnsatz}. The exponents are $\eta \simeq -0.1$, $\nu \simeq 1.0$ with $T_K$ as shown in the inset.}
\end{center}
\label{Qmoments}
\end{figure*}

In left panel of Fig. \ref{Qmoments}, the temperature dependence of the overlap ( see Method section for detail definition ) for 
different system sizes are shown and one can clearly see that for larger system sizes the overlap seems to change more sharply. 
The corresponding susceptibility is given by $\chi_{SG} = (N - N_{imp})[\langle \delta q^2 \rangle]$, where $\langle . \rangle$ is the 
thermal averaging and $[.]$ means the averaging over the different realizations of the disorder. $N_{imp}$ is the number of impurity 
particles in the system which is given by $N_{imp} = \rho_{imp}N$. In middle panel of Fig.\ref{Qmoments}, the susceptibility shows 
dramatic variation with temperature and at lower temperature the susceptibility seems to change quite strongly with system size also. 
For system size $N = 250$ the susceptibility changes almost by a factor of $50$ compare to its high temperature value. With this data divergence 
of the susceptibility at lower temperature in the thermodynamic limit can not be established but the strong increase of its value with 
the system size is encouraging. we have tried to collapse the susceptibility data for different system sizes using the 
finite size scaling ansatz ($L\equiv N^{1/3}$):
\begin{equation}
\chi_{SG} = L^{2-\eta} {\cal F}\left(L^{1/\nu}|T - T_K| \right),
\label{chiAnsatz}
\end{equation}
with rather poor results. A better collapse is shown in the right panel of Fig. \ref{Qmoments}
with $\eta \simeq -0.1$, $\nu \simeq 1.0$ and $T_K(L)$ now being a function of system size $L$. It is important to notice that as $T_K$ 
grows with system size it is reasonable that the divergence temperature will not go to zero in the thermodynamic limit. The 
data seems to indicate a possible non-zero temperature thermodynamical 
transition of second order in nature. The value of $\nu$ is somewhat larger than the exponent $\nu = 2/d$ argued in 
RFOT theory \cite{RFOT, RFOT1} by scaling arguments. These results are very similar to the one obtained for finite range $p$-spin 
glass model in 3 dimension in \cite{98CCP}. The inability to collapse the data with a size independent critical temperature may be the effect of   
the strong finite size scaling corrections: the system sizes studied here are relatively small. A different scenario would be (in the thermodynamic limit) a strong increase of the susceptibility approaching the mode coupling transition temperature followed by a cross over to a different behavior at low temperature. We noted that the susceptibility data for $N=4000$, $T>=0.45$ (that should have small finite size effects) are well fitted by a power law, with a critical temperature around $0.4$, i.e. the putative mode coupling transition. Much more extensive simulations are needed to better understand the nature of the transition, if any.

\begin{figure*}[!h]
\begin{center}
\epsfig{file=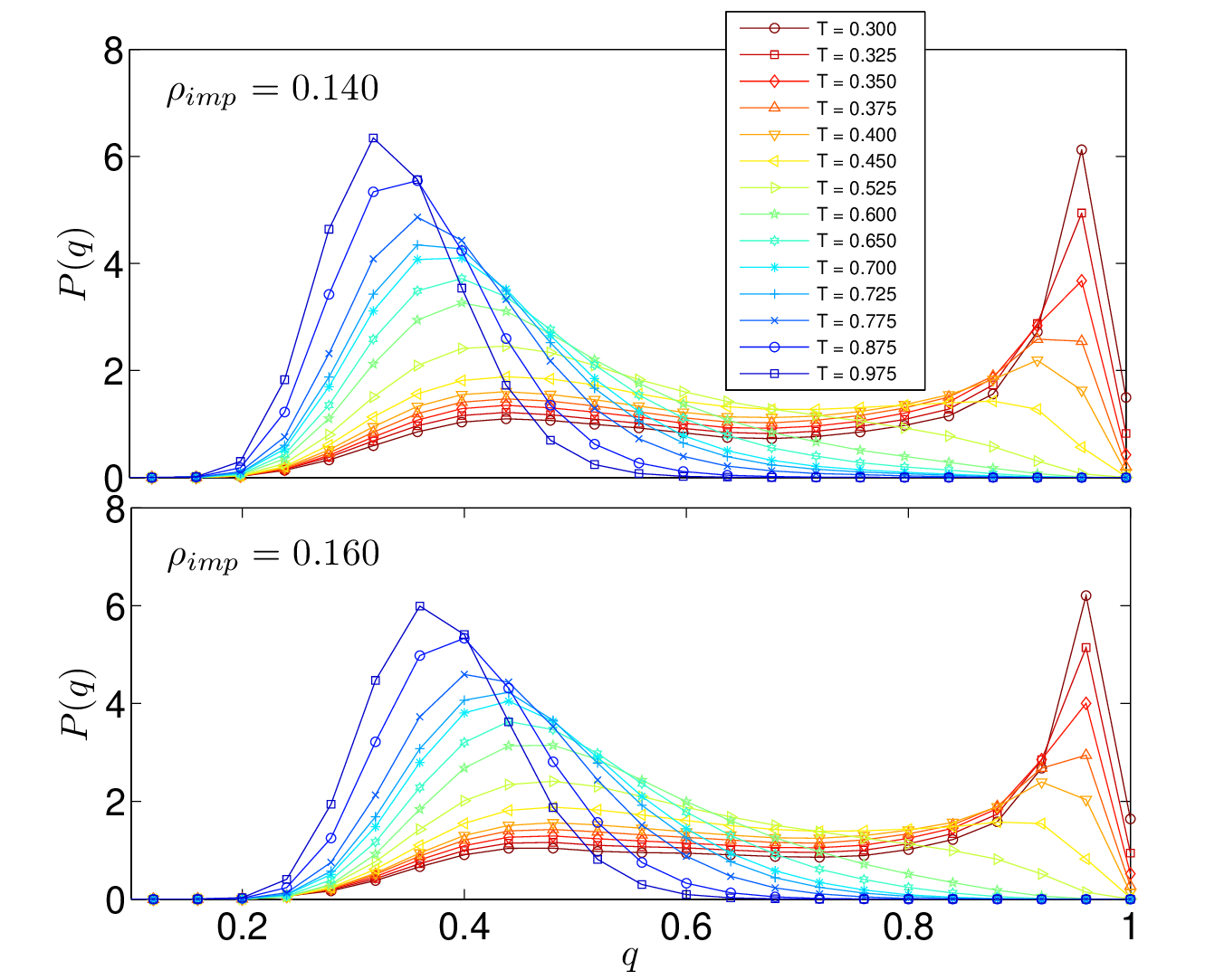,scale=0.32,angle=0,clip=} 
\epsfig{file=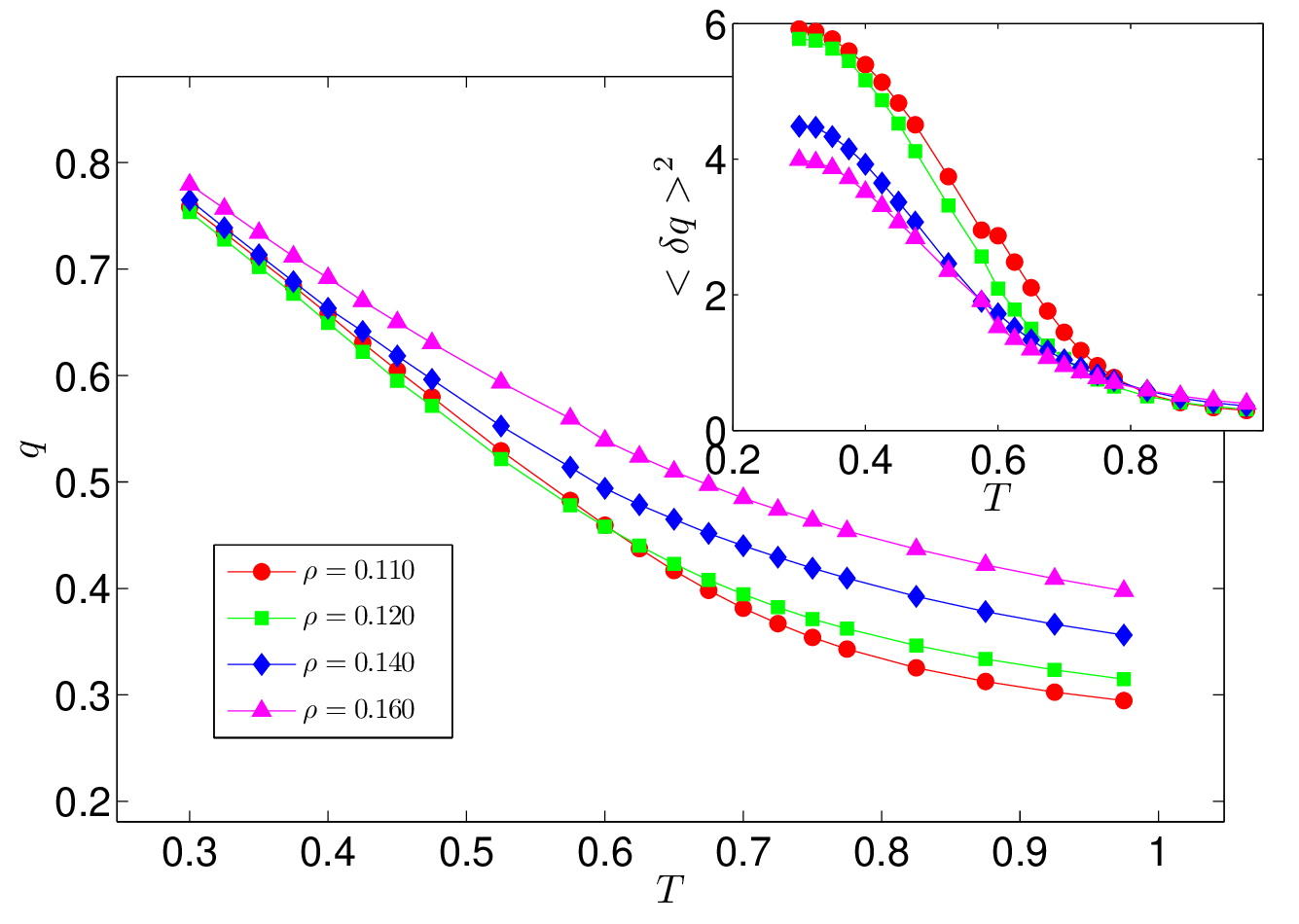,scale=0.38,angle=0,clip=} 
\caption{ Data for $N = 100$ : Left Panel : The distribution of overlap $P(q)$ system size for two different concentration of frozen particles. 
There is  not a strong qualitatively difference between these distributions for two concentrations. Right Panel : The average overlap $<q>$
as a function of temperature for different concentration of frozen particles. Top right corner inset shows the dependence of susceptibility 
with concentration $\rho_{imp}$.}
\end{center}
\label{diffRho}
\end{figure*}

Mode Coupling Theory calculation in similar geometry for the hard sphere system in \cite{12Szamel} suggests that at some critical density
($\rho_{imp} \sim 0.15$ in \cite{12Szamel}) of the frozen particles the time dependence of the two point correlation function will change from 
being two steps to one step. So to check whether we are not very close to this critical density we performed similar studies for different 
$\rho_{imp}$ for $N = 100$ and calculated the overlap distribution $P(q)$. One can clearly see in Fig.\ref{diffRho}, that there is no 
qualitative change in the behaviour with different pinning density. In right panel we show how $<q>$ changes with $\rho_{imp}$ as a function 
of temperature and the top right corner inset show the corresponding susceptibility.    

\noindent{\bf Conclusions:\\}
To conclude we showed how a very simple particle model with random pinning can be used to explore the glassy phase at deep supercooled 
regime and also showed how this model can be used to compute different spin glass correlation functions to shed light on the relation between
spin glass transition and structural glass transition which is the basis for most of the recent theories of structural glasses. This model 
can also be used to find out the relation between different length scales as in this model the length scale is calculated directly from the 
spin glass order parameter correlation function. It would be extremely interesting to arrive to a precise determination of the phase structure of the model and to compare it with accurate numerical simulations.

\begin{materials}

\noindent{\it Simulation details :} 
The interaction potential  for ${r_{ij}} \le x_c$ is given by
\begin{equation}\label{potential}
\phi\left(r_{ij}\right) =
\frac{k}{2}\left[
\left(r_{ij}\right)^{-k} 
+ \sum_{\ell=0}^{q}  c_{2\ell} 
\left(r_{ij}\right)^{2\ell}
\right] 
\end{equation} 
while it is 0 for  ${r_{ij}} > x_c$, where $r_{ij}$ is the distance between particle $i$ and $j$,  and $x_c$ is the
 dimensionless length for which the potential will vanish continuously up to $q$ derivatives.  The coefficients 
$c_{2\ell}$ are given in \cite{12KLP}.

\noindent{\it Dynamics:} We studied the dynamics using two point overlap correlation function, $Q(t)$, defined as 
\begin{eqnarray}
Q(t)= \int d\vec{r} \rho(\vec{r},t_0)\rho(\vec{r},t+t_0) 
\sim \sum_{i,j=1}^{N}w(|\vec{r}_i(t_0)-\vec{r}_j(t_0+t)|)
\label{overlapDef}
\end{eqnarray} 
where $\rho(\vec{r},t)$ is particle density at space point $\vec{r}$ at time $t$, and $w(r) = 1$, 
if $r \le a$ and zero otherwise. Average over the time origin $t_0$ is assumed. The use of the window function 
[$a = 0.30$] is to remove the fluctuations in particle positions 
due to small amplitude vibrational motion. In Eq.\ref{overlapDef} the contribution only due to the self-term is denoted as $Q_s(t)$. 
The structural relaxation time $\tau_\alpha$ is the time where $Q_s(\tau_\alpha) = 1/e$.

\noindent{\it Replica overlap and Extraction of static length scale:} 
In this section we will explain how we have extracted the growing static length scale in this system. 
We followed the method used in \cite{98P}. The method is explained briefly below, we start with the following definition 
of overlap between two replica
\begin{equation}
q = \frac{1}{N}\sum_{i=1}^{N}q_i\,\, \mbox{where}\,\, q_i = \sum_{k=i}^{N}w(x_i - x_k),\,\, 
\end{equation}
with $w(x) = 1 \,\,\mbox{for}\,\, x<0.3\,\, \mbox{else}\,\, 0,$. 
Now we define the coarse grain variable as 
\begin{equation}
\mu(x) = \sum_i \delta(x - x_i) q_i, \quad \mbox{and defined}\quad f(r) = \langle \mu(x)\mu(y) \rangle 
\end{equation}
where $r = |x-y|$. Now to remove the natural oscillation in the function $f(r)$ we divide the function by pair correlation function $g(r)$ 
to define another function $c(r) = f(r)/g(r)$, and fitted the function to the following fitting function 
\begin{equation}
\tilde{c}(x) = a + \exp(-x/\xi)\left[b + c\cos(xd+e)\right]
\label{fitting}
\end{equation}
to extract the length scale $\xi$. We extracted the correlation length scale for the $N = 512, 2000$ and $4000$ system sizes which have 
been equilibrated using standard molecular dynamics simulations using Berendsen thermostat \cite{84BPGDH} and we studied the system in the 
temperature window $T \in \{0.400, 1.000\}$. We averaged the data over $20$ different realizations of the disorder. For last three temperatures 
$T = 0.400, 0.420$ and $0.450$, we averaged the data over $40$ realizations of the disorder.

\noindent{ \it Relation between length scale and relaxation time :}
We start with the ansatz for the relation between relaxation time $\tau_{\alpha}$ and the static correlation length $\xi$.
\begin{equation}
\tau_{\alpha}(T) \propto \exp\left[ \frac{\Delta_0 \xi(T)^{\psi}}{T} \right],
\end{equation}
where $\Delta_0$ is a non-universal coefficient that depends on the details of the glass former. Now at reference temperature 
$T = T_0$ (the highest temperature in this case) we define the typical length to be $\xi_0$. Then the relaxation time at that 
temperature is
\begin{equation}
\tau_\alpha(T_0) \propto \exp\left[ \frac{\Delta_0}{T_0}\right],
\end{equation}
So we have the following relation
\begin{equation}
\log\left[ \frac{\tau_\alpha(T)}{\tau_\alpha(T_0)} \right] = \frac{\Delta_0 \xi^{\psi}}{T} - \frac{\Delta_0 \xi_0^{\psi}}{T_0}
=  \frac{\Delta_0 \xi_0^{\psi}}{T_0} \left[ \frac{{(\xi/\xi_0)}^{\psi}}{T/T_0} - 1 \right]\ .
\label{ansatz}
\end{equation}
As the pre-factor $\Delta_0$ is not known a-priori for different models we choose $\Delta_0 = 1.0, 1.11$ for $\rho_{imp} = 0.090$ and $0.110$ 
respectively in Fig.\ref{dynamicalDetails}. We also choose $\psi = 1.0$ for these two cases.

\noindent{\it Parallel tempering methods :}
To equilibrate the system still at lower temperature, we have implemented parallel tempering method. We briefly mention the method here 
as details can be found in \cite{00YK}. We construct a system consisting of $M$ non-interacting subsystems (replicas), each consists of
$N$ particles with phase space coordinate $\{P^{i},Q^{i}\}$, where $P^i = \{p_1, p_2 \ldots p_N\}$ and $Q^i = \{q_1, q_2, \ldots q_N\}$ for 
the $i^{th}$ subsystem. The Hamiltonian of the $i^{th}$ subsystem is given by
\begin{equation}
H_i(P^i, Q^i) = K(P^i) + \Lambda_i E(Q^i),
\end{equation}  
where $K$ is the kinetic energy, $E$ is the potential energy of the system and $\Lambda_i \in \{\lambda_1, \ldots \lambda_M \}$ is the 
parameter to rescale the potential energy. Now we perform molecular dynamics simulation of the whole system with Hamiltonian 
${\cal H} = \sum_i^{M}H_i$ at a reduced temperature $T = 1/\beta_0$ using a modified isokinetic simulation method. Time step for the MD
is taken to be $\delta t = 0.005$. Now at each time interval of $\Delta t_{X} = 1000 \delta t$ we exchange the parameter $\lambda$ between 
different replica $i$ and $j$ keeping all other 
things unchanged. The exchange is accepted using a Metropolis scheme, with a probability
\begin{equation}
w_{i,j} =\min( 1, 
\exp(-\Delta_{i,j}))
\end{equation}
where $\Delta_{i,j} = \beta_0 ( \Lambda_j - \Lambda_i ) (E(Q^i) - E(Q^j))$. We perform these steps for sufficiently long time such that we 
get proper equilibration of the system. This way we generate canonical distribution at inverse temperatures $\beta_i = \lambda_i\beta_0$.
Here we have also parallelized the code using MPI to speed up the simulation process. We have used $M = 12$ replicas 
between temperature $0.300$ to $0.600$ and another $12$ replicas for temperature range $0.600$ to $1.000$ for $N = 100$ and $150$ system size. 
For $N = 200$, we have used $16$ replicas for $0.300$ to $0.600$ temperature range and $12$ between $0.600$ to $1.000$. For $N = 250$, we 
only tried to equilibrate the system up to $T = 0.330$ and used $16$ replicas between temperature range $0.330$ to $0.600$ and $12$ between 
$0.600$ to $1.000$. We have averaged the data over $400$ different realizations of the disorder for systems $N = 100, 150, 200$ and $250$.
So the estimated computer time for $N = 250$ system size is close to $10^5$ hours. To check the equilibration of the system we used the 
method as in \cite{00YK} to rescale the Canonical distribution function using the formula 
\begin{equation}
P_i(E;T_j=\lambda_j\beta_0) = \frac{P_i(E)\exp\left[ (\lambda_i - \lambda_j)\beta_0E\right]}
{\int dE'P_i(E')\exp\left[ (\lambda_i - \lambda_j)\beta_0E'\right]},
\label{reweighting}
\end{equation}
In equilibrium the left hand side of the above equation should be independent of $i$ to within the accuracy of the data as can be seen in 
left panel of Fig.\ref{parllTmp}.
\end{materials}

\begin{acknowledgments}
We want to thank Prof. Grzegorz Szamel for many useful discussion during his visit to Rome. This project is supported by European Research Council  with grant Project 
No. 247328.

\end{acknowledgments}



\end{article}

\begin{thebibliography}{100}
\bibitem{09Cav}
Cavagna A (2009) {\it Supercooled liquids for pedestrians} - Phys. Rep. {\bf 476}:51-124.

\bibitem{11BB}
Berthier L and Biroli G, (2011) {\it Theoretical perspective on the glass transition and amorphous materials} - 
Rev. Mod. Phys. {\bf 83}:587-645.

\bibitem{BookParisiSpin87}
Mezard M, Parisi G, and Virasoro M, (1987) {\it SPIN GLASS THEORY AND BEYOND
An Introduction to the Replica Method and Its Applications } - World Scientific Lecture Notes in Physics - Vol. 9.

\bibitem{05CC}
Castellani T and Cavagna A, (2005) {\it Spin-glass theory for pedestrians}- J. Stat. Mech. {\bf 2005}:P05012 - P05064.

\bibitem{87KT}
Kirkpatrick TR and Thirumalai D, (1987) {\it p-spin-interaction spin-glass models: Connections with the structural glass problem} - 
Phys. Rev. B {\bf 36}:5388-5397.

\bibitem{87KW}
Kirkpatrick TR and Wolynes PG, (1987) {\it Connections between some kinetic and equilibrium theories of the glass transition} - 
Phys. Rev. A {\bf 35}:3072-3080.

\bibitem{87KWa} 
Kirkpatrick TR and Wolynes PG, (1987) {\it Stable and metastable states in mean-field Potts and structural glasses} - 
Phys. Rev. B {\bf 36}:8552 - 8564.

\bibitem{99FP}
Franz S and Parisi G, (1999) {\it Critical properties of a three-dimensional $p$-spin model} - Eur. Phys. J. B {\bf 8}: 417.

\bibitem{02MD}
Moore MA and Drossel B, (2002) {\it p-Spin Model in Finite Dimensions and Its Relation to Structural Glasses } -
Phys. Rev. Lett. {\bf 89}:217202.

\bibitem{RFOT}
Kirkpatrick TR, Thirumalai D, Wolynes PG, (1989) {\it Scaling concepts for the dynamics of viscous liquids near an ideal glassy state} - 
Phys Rev A {\bf 40}:1045.

\bibitem{BAFRPA}A. Barrat,  S. Franz and  G. Parisi  { J. Phys. A: Math. Gen.} {\ 30}, 5593 (1997). 

\bibitem{MePa1} M. M\'ezard and G. Parisi, (1999) {\em Thermodynamics of glasses: a first principles computation} J. Phys.: Condens. Matter {\bf 11 A} 157-188.

\bibitem{RFOT1}
Lubchenko V, Wolynes PG, (2007) {\it Theory of structural glasses and supercooled liquids} - Annu Rev Phys Chem {\bf 58}:235.

\bibitem{RFOT2}
Biroli G and Bouchaud J-P, (2012) in {\it Structural Glasses and Supercooled Liquids: Theory, Experiment, 
and Applications} - edited by P.G. Wolynes, V. Lubchenko, John Wiley \& Sons.
\bibitem{12MP}
M\`ezard M and Parisi G, (2012) in {\it Structural Glasses and Supercooled Liquids: Theory, Experiment, 
and Applications} - edited by P.G. Wolynes, V. Lubchenko, John Wiley \& Sons.
\bibitem{65AG}
Adam G, Gibbs JH (1965) {\it On the temperature dependence of cooperative relaxation properties in glass-forming liquids}. 
J Chem Phys {\bf 43}:139–146.

\bibitem{98CCP}
Campellone M, Coluzzi B and Parisi G, (1998) {\it Numerical study of a short-range p-spin glass model in three dimensions}, 
Phys. Rev. B {\bf 58}, 12081–12089.

\bibitem{88CAA}
Angel CA, (1988) {\it Perspective on the glass transition} - J. Phys. Chem. Solids {\bf 49}:863–871.

\bibitem{92GS}
G\"otze W, ~Sj\"ogren L (1992) {\it Relaxation processes in supercooled liquids}. Rep Prog Phys {\bf 55}:241–376.

\bibitem{00Ediger}
Ediger MD (2000) {\it Spatially heterogeneous dynamics in supercooled liquids}. Annu Rev Phys Chem {\bf 51}:99–128.

\bibitem{05Berthier}
Berthier L, et al. (2005) {\it Direct experimental evidence of a growing length scale accompanying the glass transition}. 
Science {\bf 310}:1797–1800.

\bibitem{98P} G. Parisi {\it An increasing correlation length in off-equilibrium glasses}J. Phys. Chem. B, 1999 {\bf 1999}, pp 4128Ð4131.

\bibitem{06BBMR}
Biroli G, Bouchaud J-P, Miyazaki K, Reichman D R (2006) {\it Inhomogeneous mode-coupling theory and growing dynamic length 
in supercooled liquids}. Phys Rev Lett {\bf 97}:195701–1–195701–4.

\bibitem{09KDS}
Karmakar S, Dasgupta C, Sastry S (2009) {\it Growing length and time scales in glass forming liquids}. Proc Nat. Acad Sci USA 
{\bf 106}:3675â3679.

\bibitem{08BBCGV}
Biroli G, Bouchaud J-P, Cavagna A, Grigera T S, Verrocchio P (2008) {\it Thermodynamic signature of growing amorphous order in 
glass-forming liquids}. Nat Phys {\bf 4}:771–775.

\bibitem{09KL}
Kurchan J and Levine D (2009) {\it Correlation length for amorphous systems}. arXiv:0904.4850.

\bibitem{10MGIO}
Mosayebi M, Del Gado E, Ilg P, and \"Ottinger HC (2010) {\it Probing a Critical Length Scale at the Glass Transition}. 
Phys. Rev. Lett. {\bf 104}:205704-1 - 205704-4.

\bibitem{12KLP}
Karmakar S, Lerner E, and Procaccia I, (2012) {\em Direct estimate of the static length-scale accompanying the glass transition} 
Physica A {\bf 391}:1001.

\bibitem{12CB}
Cammarota C and Biroli G, (2012) {\it Ideal Glass Transitions by Random Pinning}, Proc. Nat'l. Acad. Sci. USA {\bf 109}:8850-8855.

\bibitem{11Krakoviack}
Krakoviack V (2011) {\it Mode-coupling theory predictions for the dynamical transitions of partly pinned fluid systems}, 
Phys. Rev. E {\bf 84}, 050501(R).

\bibitem{12Szamel}
Szamel G and Flenner E (2012), {\it Glassy dynamics of partially pinned fluids: an alternative mode-coupling approach}, arXiv:1204.6300.

\bibitem{03Kim}
Kim K (2003) {\it Effects of pinned particles on the structural relaxation of supercooled liquids}. Europhys Lett {\bf 61}:790–795.

\bibitem{11KMS}
Kim K, Miyazaki K, Saito S (2011) {\it Slow dynamics, dynamic heterogeneities, and fragility of supercooled liquids confined in 
random media}. J Phys: Condens Mat {\bf 23}:234123.

\bibitem{12BK}
Berthier L, Kob W (2012) {\it Static point-to-set correlations in glass-forming liquids}. Phys Rev E {\bf 85}, pp 011102-1–011102-5.

\bibitem{11CIPZ}
Charbonneau P, Ikeda A, Parisi G, and Zamponi F (2011) {\it Glass Transition and Random Close Packing above Three Dimensions}. 
Phys. Rev. Lett. {\bf 107}:185702-1 - 185702-4.

\bibitem{12HMR}
Hocky GM, Markland TE, and Reichman DR (2012) {\it Growing Point-to-Set Length Scale Correlates with Growing Relaxation Times 
in Model Supercooled Liquids}. Phys. Rev. Lett. {\bf 108}:225506-1 - 225506-5. 

\bibitem{JANUS} F Belletti, A Cruz, LA Fernandez, A Gordillo-Guerrero, M Guidetti, A Maiorano, F Mantovani, E Marinari, V Martin-Mayor, J Monforte, A Mu–oz Sudupe, D Navarro, G Parisi, S Perez-Gaviro, JJ Ruiz-Lorenzo, SF Schifano, D Sciretti, A Tarancon, R Tripiccione, D Yllanes (2009) {\it An In-Depth View of the Microscopic Dynamics of Ising Spin Glasses at Fixed Temperature}  J. of Stat. Phys.
{\bf 135} 1121-1158.

\bibitem{12KP}
Karmakar S, Procaccia I (2012) {\it Finite Size Scaling for the Glass Transition: the Role of a Static Length Scale}. arXiv:1204.6634.

\bibitem{98CoPa}B. Coluzzi and G. Parisi  (1998) {\it On the Approach to the Equilibrium and the Equilibrium Properties of a Glass-Forming Model}. J. Phys. A: Math. Gen. {\bf 31} 4349-4362.

\bibitem{00YK}
Yamamoto R and Kob W (2000) {\em Replica-exchange molecular dynamics simulation for supercooled liquids} Phy.Rev. E {\bf 61}:5473.


\bibitem{84BPGDH}
Berendsen HJC, Postma JPM, van Gunsteren WF, Dinola A, and Haak JR, (1984) {\it Molecular dynamics with coupling to an external bath }.
J. Chem. Phys. {\bf 81}, 3684-3690.

\end{thebibliography}
\end{document}


\title{Supplementary Material : Random Pinning Glass Model}

\author{Smarajit Karmakar and Giorgio Parisi}
\affiliation{Departimento di Fisica, Universit\'a di Roma ``La Sapienza '',INFN,Sezione di Roma I, IPFC – CNR,
Piazzale Aldo Moro 2, I-00185, Roma, Italy}
\maketitle

\section{Time-Temperature Superposition for $Q(t)$ : }
\begin{figure}[!h]
\begin{center}
\epsfig{file=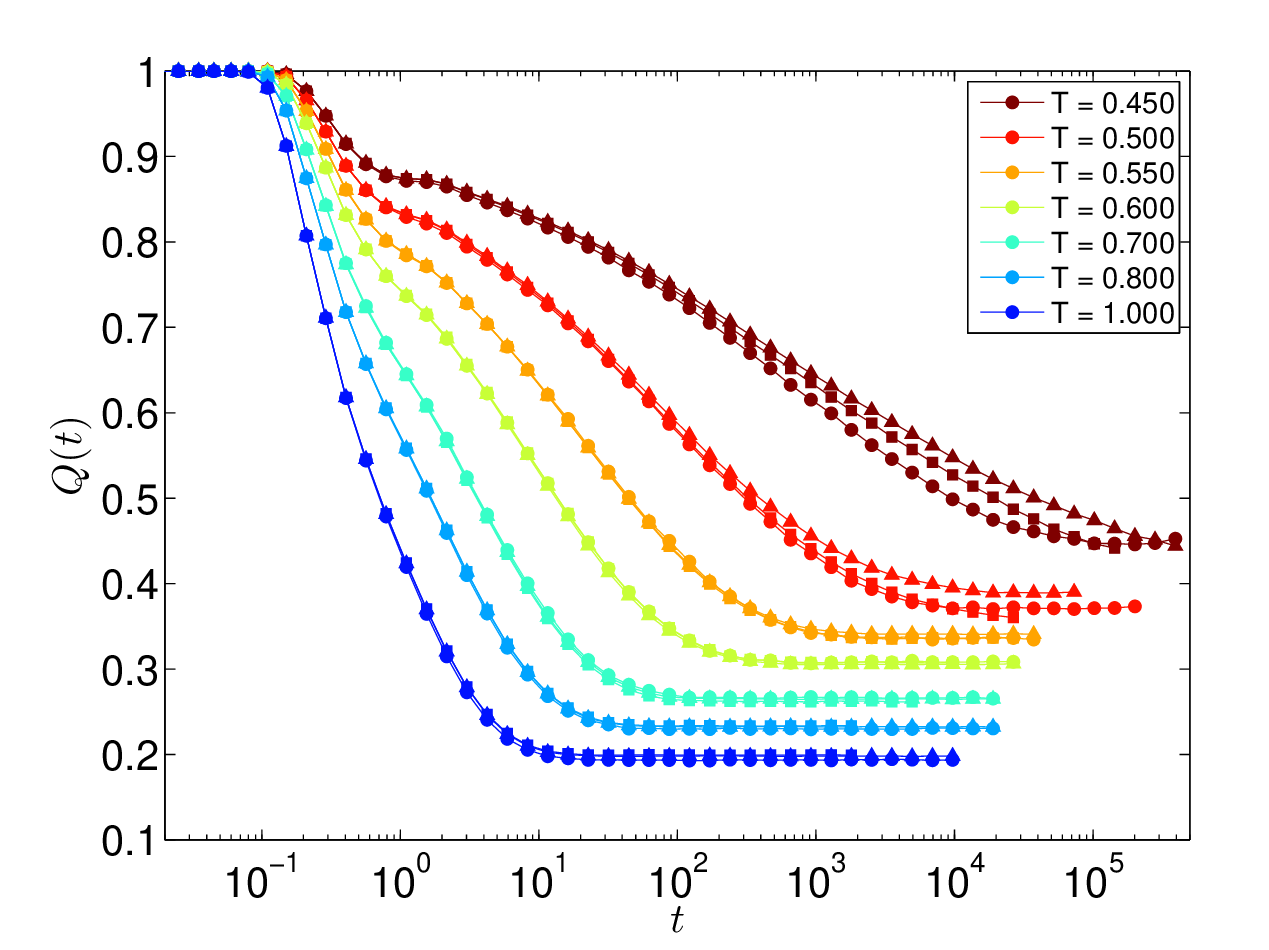,scale=0.40,angle=0,clip=} 
\epsfig{file=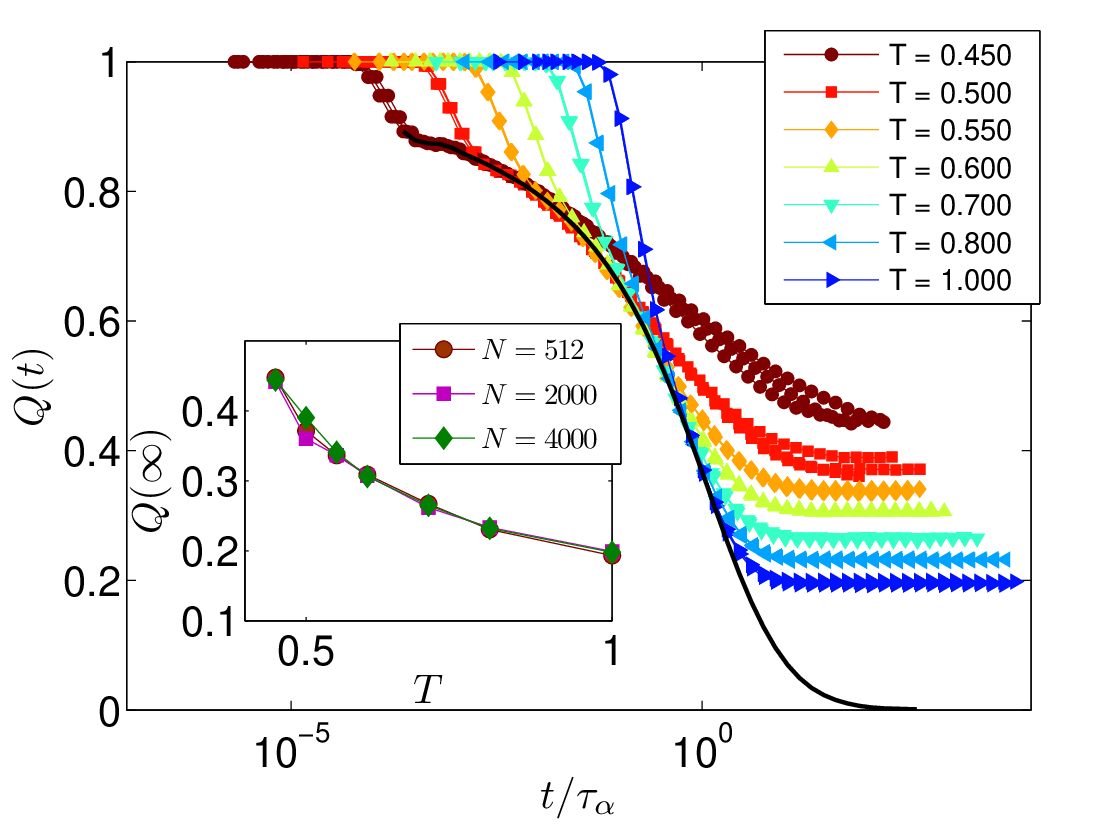,scale=0.46,angle=0,clip=}
\caption{Data for $N=512$, $N=2000$ and $N=4000$. Left Panel: System size dependence of the relaxation function $Q(t)$. One can see that there is very small finite 
size effect in these system sizes. Right Panel: The time-temperature superposition of $Q(t)$ using the relaxation time
$\tau_{\alpha}$ obtained from the self part of $Q(t)$ from the condition $Q_s(\tau_{\alpha}) = 1/e$. It clearly shows
that the initial part of the relaxation function is very much same as that of the self part but deviates from it as the 
infinite time value of $Q(t)$ is not $0$ in this case.  The black solid line is the master curve one would obtain if one 
tries to do the time-temperature superposition for $Q_s(t)$ as done in Fig.[1] in main article. This is shown just to point 
out that the master curve for $Q(t)$ also follows the same curve but deviates from it due to non-zero asymptotic value. 
Inset shows the temperature dependence of the infinite time value of $Q(t)$ that is $Q(\infty)$ for all the system sizes. 
One can cleary see that there is hardly any system size dependence in this quantity.}
\label{timeTempSup}
\end{center}
\end{figure}
In SMFig.\ref{timeTempSup}, we have shown the time-temperature superposition for the full overlap correlation function $Q(t)$ 
with the same relaxation time $\tau_{\alpha}$ which are used to collapse the data for self part of the correlation function, 
i.e. $Q_s(t)$. The figure clearly shows that initial decay of the correlation function is same as that of the self part only. 
$Q(t)$ deviates from the master curve because of the non zero asymptotic value $Q(\infty)$. In this figure we also plotted 
the data for all system sizes. Different color indicates the temperatures and symbols distinguishes the different system sizes.
The inset shows the temperature dependence of the asymptotic value of $Q(t)$ i.e. $Q(\infty)$ and one can clearly see that 
there is very small finite size effects in this quantity. 

\section{Extraction of Static Length Scale from the Replica Correlation Function :}
\begin{figure}[!h]
\begin{center}
\epsfig{file=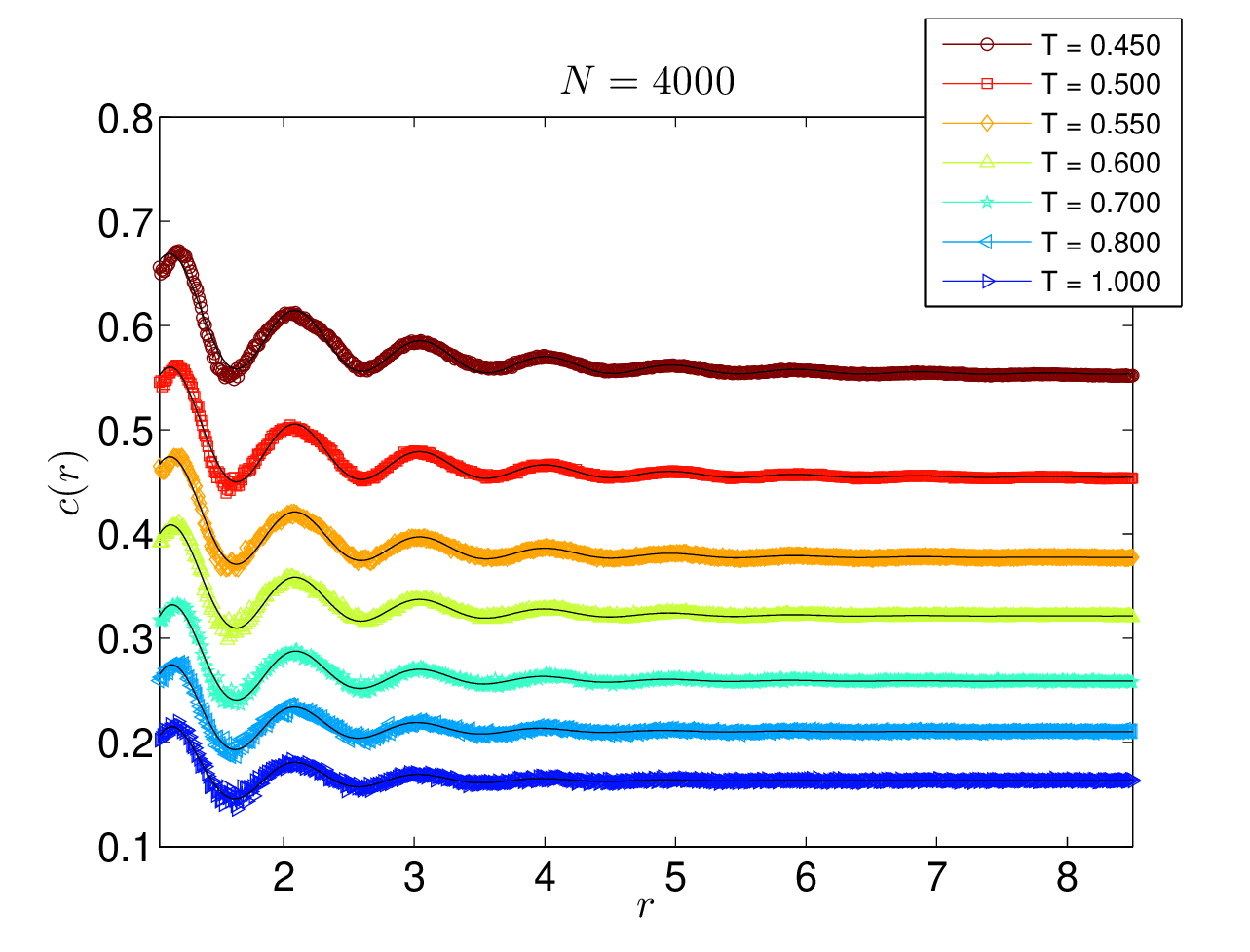,scale=0.50,angle=0,clip=} 
\caption{Extraction of the static length scale for $N = 4000$ system size. We used the functional form 
$\tilde{c}(x) = a + \exp(-x/\xi)\left[ b + c \cos(xd + e) \right]$ to fit the data. The nice fitting over the whole range of the data
indicates that the length scale extracted this way is very reliable. Data for different curve is shifted vertically by 0.01 
from each other for clarity.}
\label{crFitting}
\end{center}
\end{figure}
As discussed in the method section of the main paper about how we extracted the length scale scale from the replica correlation 
function, we have presented here some data for the $N = 4000$ system size in SMFig.\ref{crFitting}. We used the method explained 
in the Replica Overlap and Extraction of static length scale section of the Method section of the main paper. The nice fitting 
of the data over the whole range and for all temperatures indicates that the extracted length scale is very reliable. This can 
also be seen from the data shown in Fig.[1] of the main paper where we compared the results of this fitting for different system
sizes. The data for $N = 2000$ is very close to the one obtained from $N = 4000$ system size. This implies that this method of
extraction of the length scale is very robust once we rule out the finite size effect in the data.

\section{Check of Equilibration in Parallel Tempering Runs:}
\begin{figure}[!h]
\begin{center}
\epsfig{file=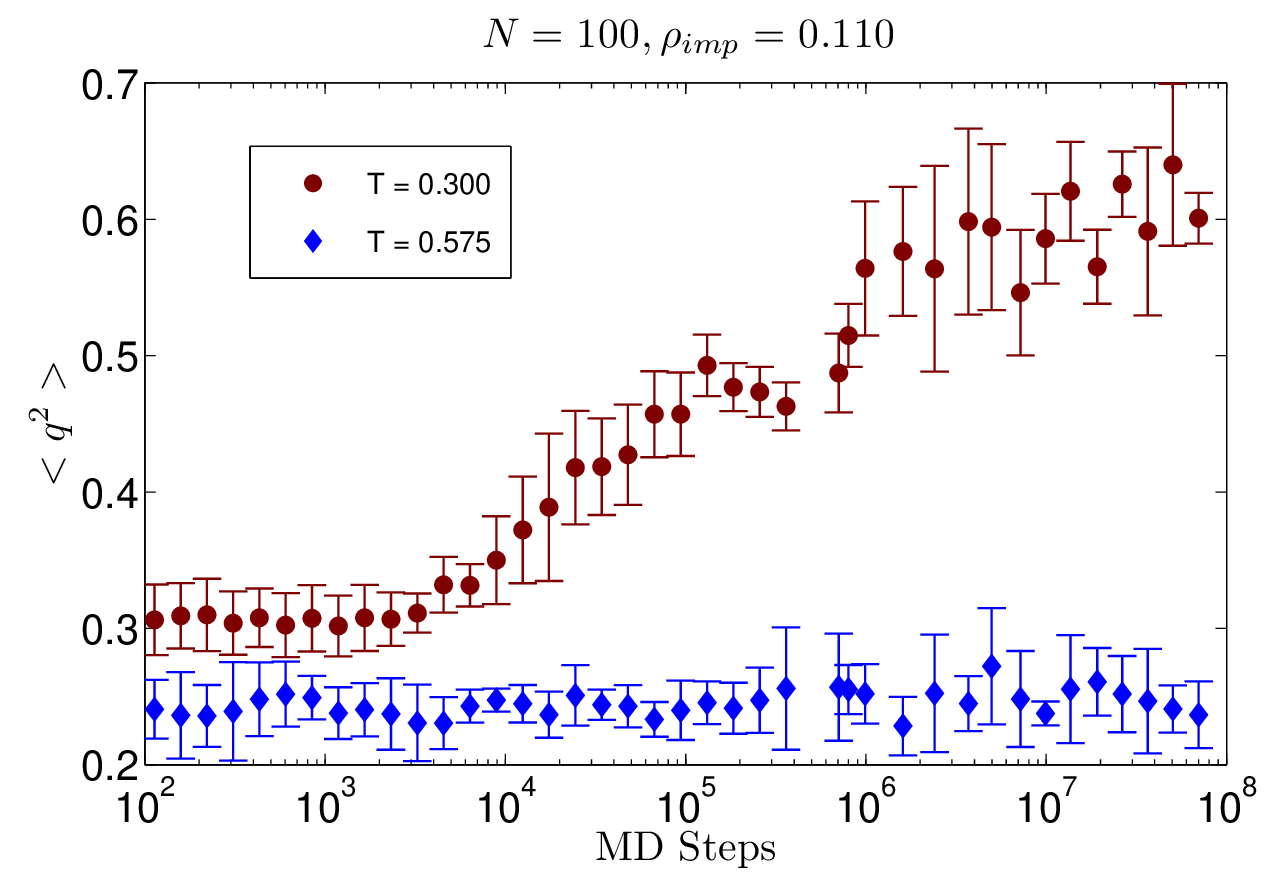,scale=0.45,angle=0,clip=} 
\caption{Evolution of $<q^2>$ as a function of MD steps to check whether the time evolution of $<q^2>$ reached a plateau. 
We take the sample to be equilibrated if $<q^2>$ reaches a plateau value within the error bar.}
\label{evolutionQsq}
\end{center}
\end{figure}

We performed two separate tests to check the equilibration of our sample : one is on the distribution of energy $P(E)$ where we 
tried to rescale all the $P(E)$ for different temperatures to collapse on the distribution of our reference temperature $T_0$ as 
shown in the Fig.[2] of the main article using the ansatz Eq.[12] in Parallel tempering part of the Method section. Here we present 
the data for 
the second test. The test consists of checking when square of the replica overlap $<q^2>$ reaches a steady plateau value with time. 
In SMFig.\ref{evolutionQsq}, we showed data for $N = 100$ system size and the data clearly shows that our simulation time 
is longer than the time required to reach a plateau value for $<q^2>$. As we have started from two completely uncorrelated replicas
their initial overlap is very small and over the time it reaches the equilibrium value from below. 
\begin{figure}[!h]
\begin{center}
\epsfig{file=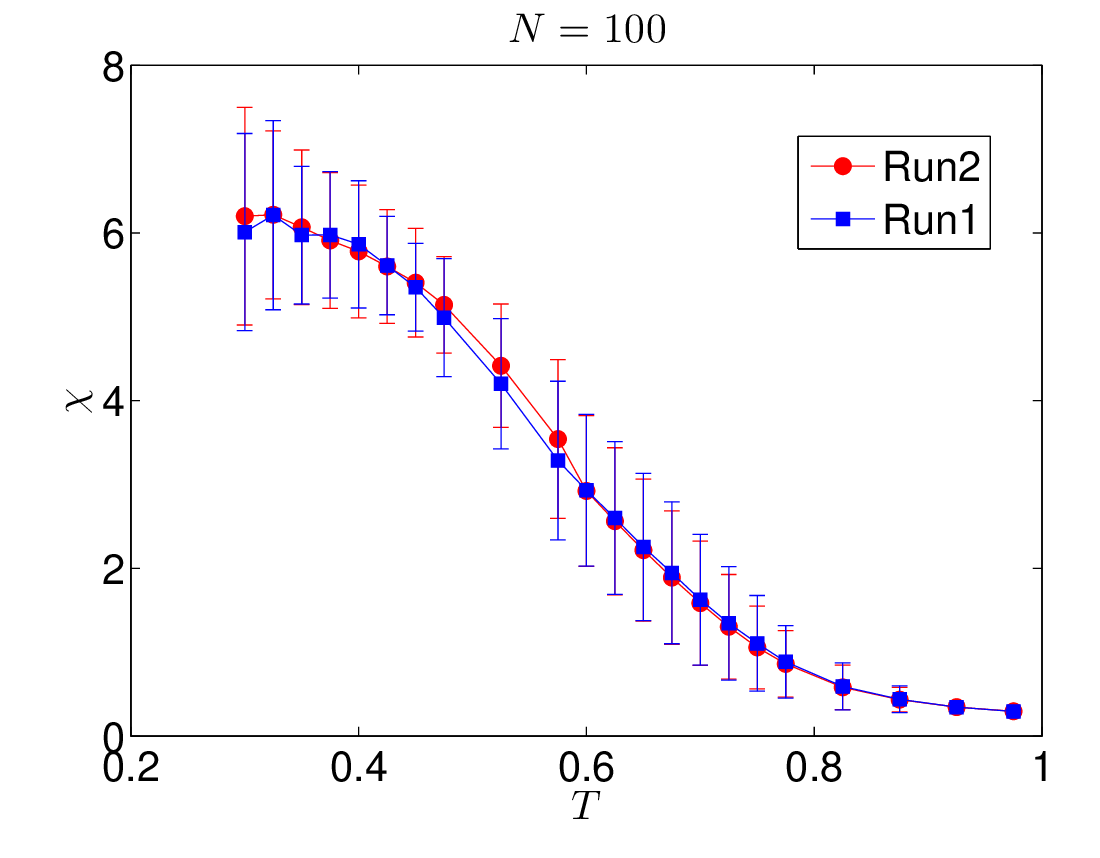,scale=0.42,angle=0,clip=} 
\epsfig{file=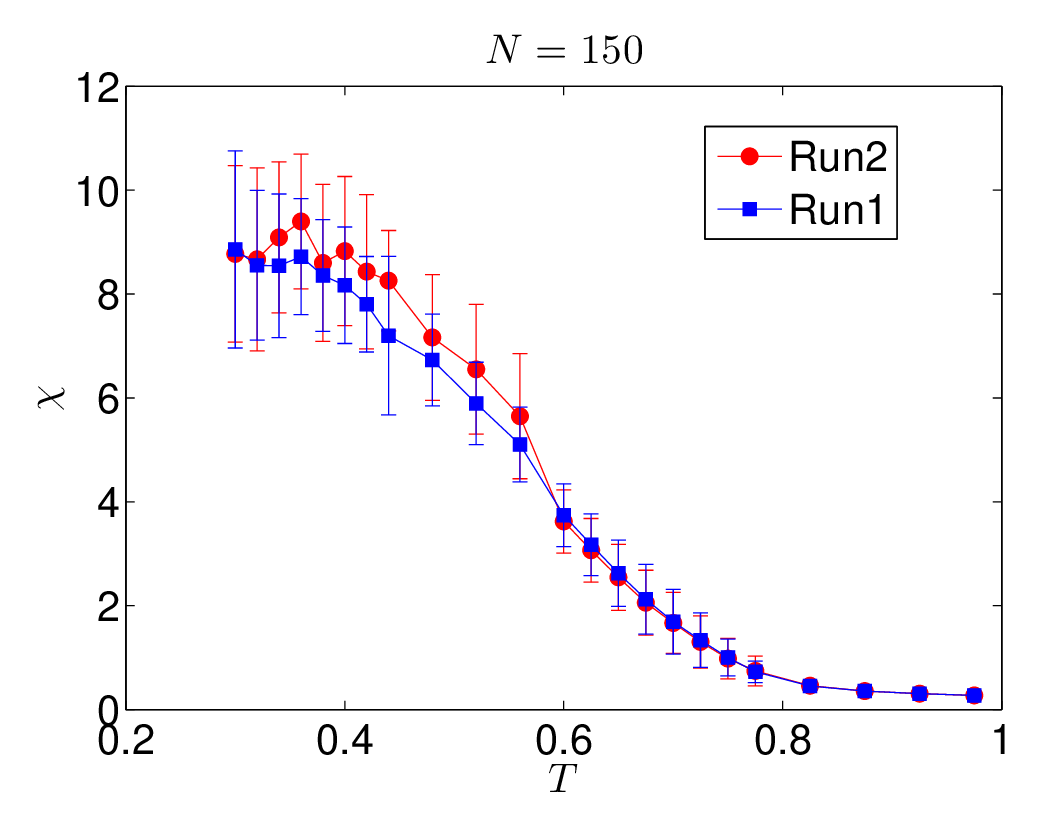,scale=0.44,angle=0,clip=}
\epsfig{file=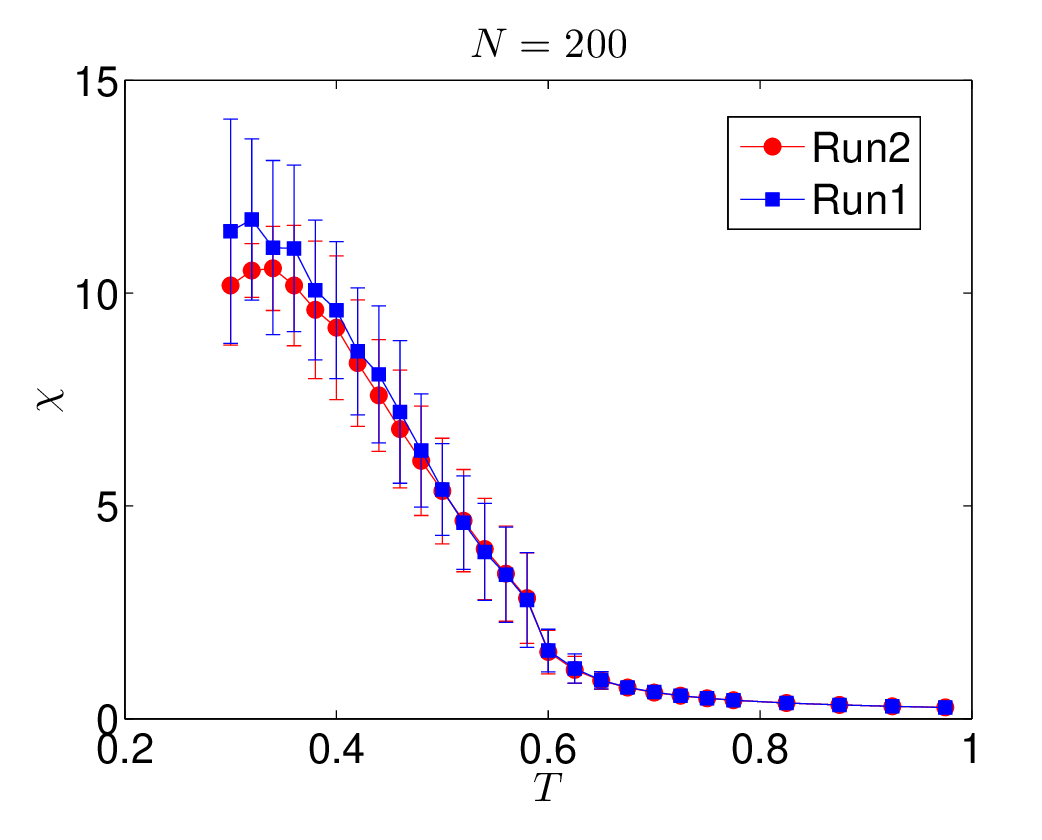,scale=0.44,angle=0,clip=}
\epsfig{file=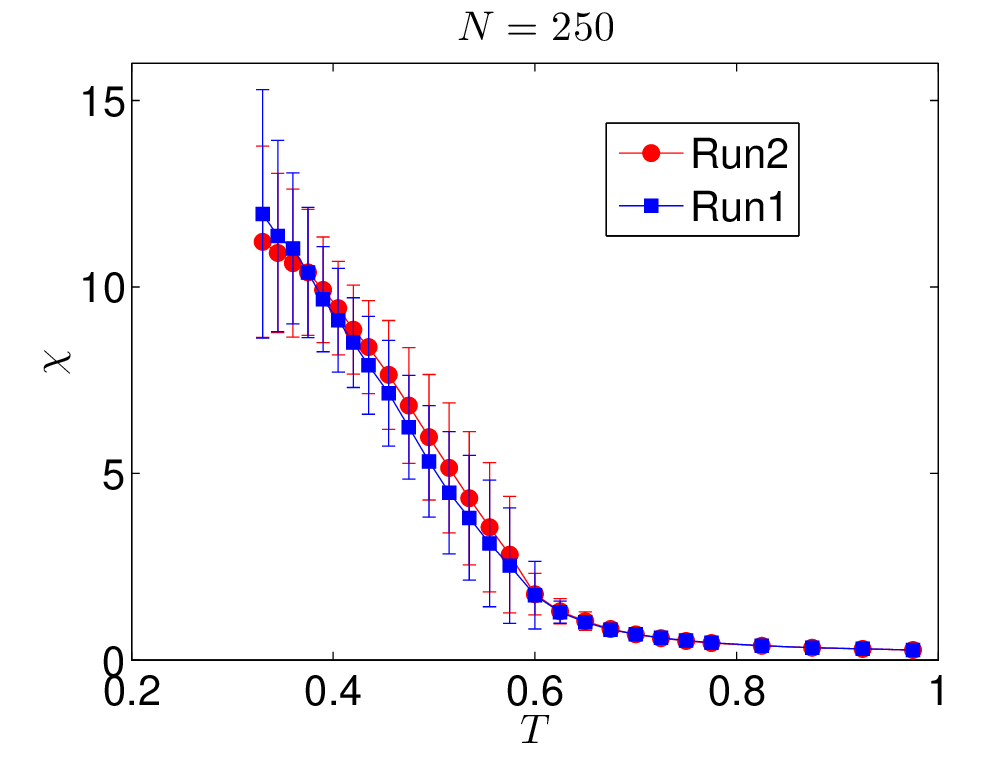,scale=0.48,angle=0,clip=}
\caption{Comparison of susceptibility $\chi$ for two sets of runs for different system sizes. Run2 is 2
 times longer than Run1. Error bars are calculated by bootstrap method.}
\label{comparison}
\end{center}
\end{figure}

In SMFig.\ref{comparison}, we have shown the spin glass susceptibility $\chi$ calculated for all the 4 different system sizes for two different
run length. Run1 is $1\times 10^8$ MD steps long and the Run2 is 2 times longer than the Run1. The value of the susceptibility is within error bar
of each other. This also confirms that our data is well equilibrated.

\section{Sample to Sample variation of $P(q)$ : }
\begin{figure}[!h]
\begin{center}
\epsfig{file=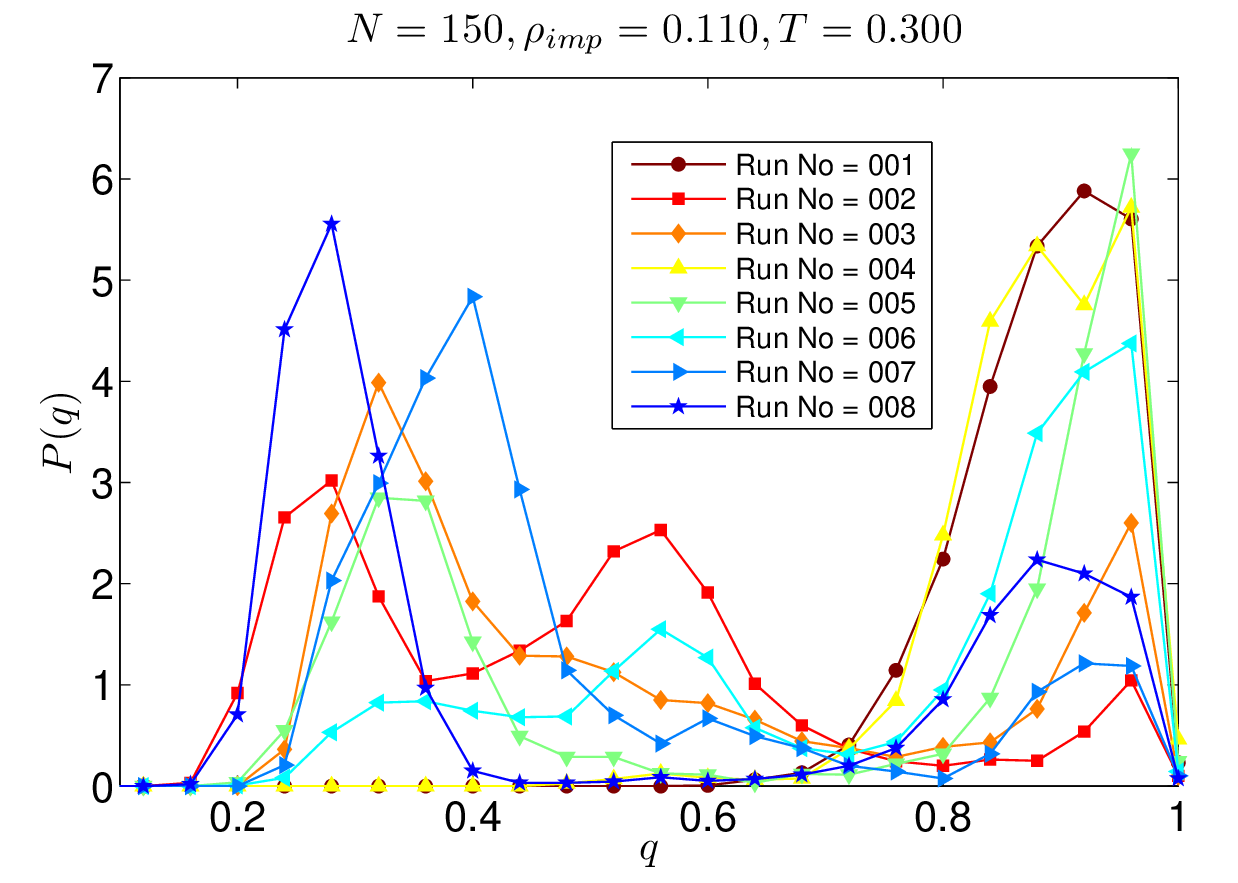,scale=0.35,angle=0,clip=} 
\epsfig{file=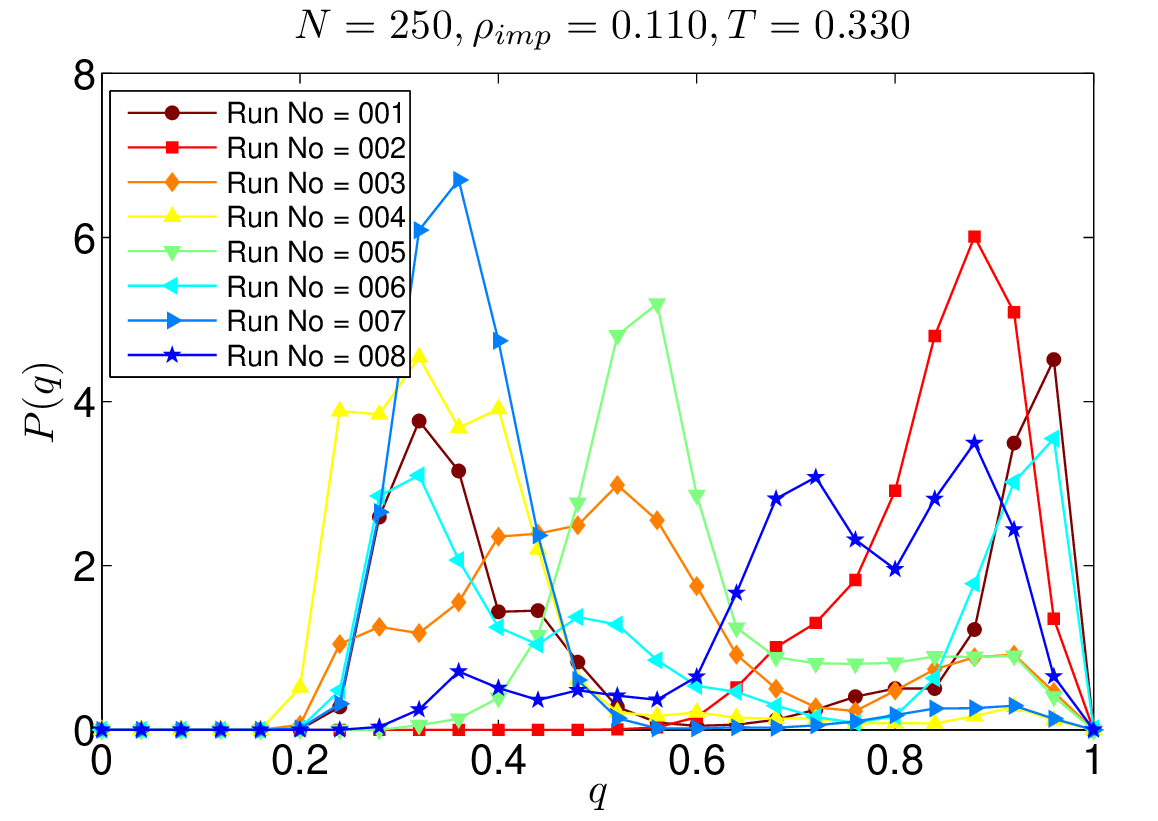,scale=0.38,angle=0,clip=}
\caption{Left Panel : Distribution of overlap $P(q)$ for $N = 150$ system size with $\rho_{imp} = 0.110$ for the temperature
$T = 0.300$ for 8 different samples. Right Panel : Distribution for the $N = 250$ system size at $T = 0.330$.}
\label{sampleToSampleFluc}
\end{center}
\end{figure}

In this section we have shown the sample to sample fluctuations of the distribution of $P(q)$ for two different system sizes at 
the lowest temperature simulated for two system sizes $N = 150$ and $250$. The SMFig. \ref{sampleToSampleFluc} shows $P(q)$ for 
$8$ randomly chosen different samples. One can clearly see the strong sample to sample fluctuations of this distribution which is also very 
similar to spin glass models \cite{12PT}. Some samples have only two peaks, with a deep valley in between, while other samples have a third peak.

\section{Skewness and Kurtosis:}
\begin{figure}[!h]
\begin{center}
\epsfig{file=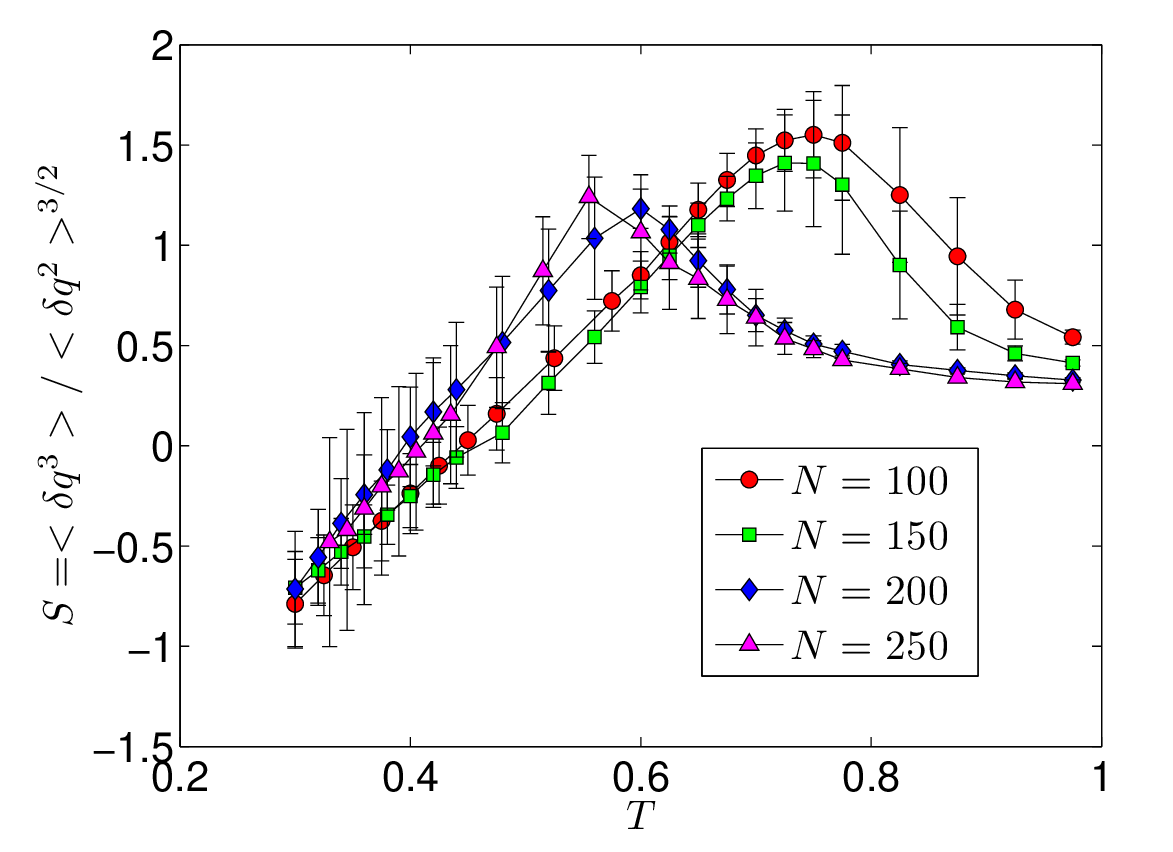,scale=0.39,angle=0,clip=} 
\epsfig{file=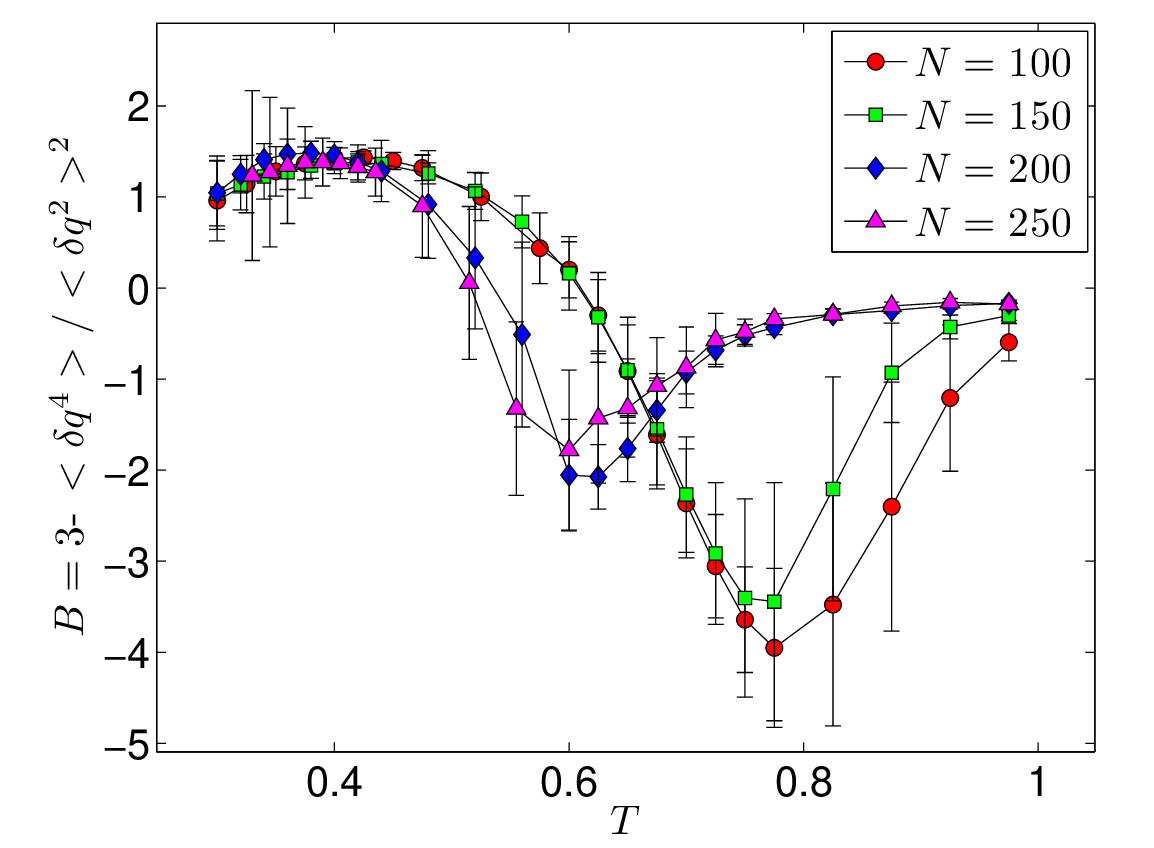,scale=0.38,angle=0,clip=}
\caption{Left Panel : Temperature dependence of skewness $S$ for different system sizes with $\rho_{imp} = 0.110$. Right Panel : 
Kurtosis for the same data set.}
\label{QskewKurt}
\end{center}
\end{figure}
In Fig. \ref{QskewKurt}, we have shown the skewness ($S$) and the kurtosis ($B$) defined as
\begin{equation}    
S = \frac{[\langle \delta q^3 \rangle]}{[\langle \delta q^2 \rangle]^{3/2}}, \quad\quad
B = 3.0 - \frac{[\langle \delta q^4 \rangle]}{[\langle \delta q^2 \rangle]^{2}} \quad \mbox{with} \quad \delta q = q - <q>. 
\end{equation}
The usual Binder cumulant is equal to $g(T) = B/2.0$ according to our definition.
Notice that $g(T)$ is non-monotonic function of temperature. In conventional phase transition this is very useful 
quantity which has been used extensively in finite size scaling analysis for determining the critical temperature in 
the thermodynamic limit. This quantity for different system sizes cross each other at critical temperature because of the 
finite size scaling. 

At zero magnetic field in the Sherrington-Kirkpatrick model for spin glass  and  the short range Edward-Anderson model $g(T)$
is always positive function of $T$ and increases regularly with decreasing temperature and the skewness is identically zero. 
Unfortunately, as soon the magnetic field is different from zero already in the Sherrington-Kirkpatrick model (where mean field 
theory is correct) the behavior at the transition point for the cumulant is much more complex and also for $1024$ spins we are 
very far from the asymptotic limit \cite{03BiCo}.

The behaviour we see here is distinctly 
different from this simple case. Moreover it is non-monotonic and mostly a negative function of temperature over the almost 
whole range. This stems out from the fact that the distribution of $q$, $P(q)$ starts to become non-Gaussian at quite high 
temperature. This behaviour is similar to the one seen for the finite range $3$-spin model in 4 dimensions \cite{99PPR}, 
where the skewness is strongly decreasing towards the large negative values when the temperature decreases.
This similarity once again seems to suggest that this model may also in the same universality class that of 
short range $3$-spin glasses which apparently have their own universality class. 
We need further study to clearly find out the exact nature of the transition if any.